\documentclass[%
preprint,
 amsmath,amssymb,
 aps,
]{revtex4-2}

\usepackage{graphicx}
\usepackage{dcolumn}
\usepackage{bm}
\usepackage{setspace}

\usepackage{xcolor}

\begin{document}


\title{Resolvent analysis for predicting energetic structures \\in the far wake of a wind turbine}

\author{Dachuan Feng} 
\affiliation{Department of Mechanical and Aerospace Engineering, Hong Kong University of Science and Technology, Hong Kong, China}
\affiliation{Guangdong Provincial Key Laboratory of Turbulence Research and Applications, Department of Mechanics and Aerospace Engineering, Southern University of Science and Technology, Shenzhen, 518055, PR China}
	
\author{Vikrant Gupta}%
\email{vik.gupta@cantab.net}
\affiliation{Guangdong Provincial Key Laboratory of Turbulence Research and Applications, Department of Mechanics and Aerospace Engineering, Southern University of Science and Technology, Shenzhen, 518055, PR China}
\affiliation{Guangdong-Hong Kong-Macao Joint Laboratory for Data-Driven Fluid Mechanics and Engineering Applications, Southern University of Science and Technology, Shenzhen, 518055, PR China
}


\author{ Larry K.B. Li}
\email{larryli@ust.hk}
\affiliation{Department of Mechanical and Aerospace Engineering, Hong Kong University of Science and Technology, Hong Kong, China}
\affiliation{Guangdong-Hong Kong-Macao Joint Laboratory for Data-Driven Fluid Mechanics and Engineering Applications, Hong Kong University of Science and Technology, Hong Kong, China}

\author{Minping Wan}
\email{wanmp@sustech.edu.cn}
\affiliation{Guangdong Provincial Key Laboratory of Turbulence Research and Applications, Department of Mechanics and Aerospace Engineering, Southern University of Science and Technology, Shenzhen, 518055, PR China}
\affiliation{Guangdong-Hong Kong-Macao Joint Laboratory for Data-Driven Fluid Mechanics and Engineering Applications, Southern University of Science and Technology, Shenzhen, 518055, PR China}
\affiliation{Jiaxing Research Institute, Southern University of Science and Technology, Jiaxing, 314031, PR China}



\begin{abstract}

A thorough understanding of the energetic flow structures that form in the far wake of a wind turbine is essential for accurate turbine wake modeling and wind farm performance estimation. We use resolvent analysis to explore such flow structures for a turbine operating in a neutral atmospheric boundary layer and validate our results against data-driven modes extracted through spectral proper orthogonal decomposition. Our results confirm that convective instabilities play a dominant role in generating turbulent kinetic energy (TKE) in the far wake. Additionally, we find evidence of the non-modal Orr mechanism contributing to TKE generation, particularly at low Strouhal numbers. The resolvent analysis method requires only the mean wake velocity and eddy viscosity profiles as inputs but can capture the energetic modes and TKE spectra in the far wake. In this specific application, the resolvent analysis method approximates the wake to be axisymmetric, which suggests that it can be paired with engineering wake models. Overall this study demonstrates the use of resolvent analysis as a viable tool for estimating TKE and for uncovering the mechanism of TKE generation.
\end{abstract}

\maketitle


\section{Introduction}\label{sec-wakeRes-intro}

Wind farms consisting of arrays of wind turbines are widely used for harnessing wind energy. In such farms, turbines are typically spaced at intervals of 4--10 turbine diameters \cite{porte2020wind}. This close spacing implies that the downstream turbines routinely operate in the far wake (often defined as the downstream region further than 4 turbine diameters) of their upstream counterparts. As a result, turbulent fluctuations in the far wakes of wind turbines significantly increase the unsteady aerodynamic loads and power fluctuations in wind farms \cite{stevens2017flow, yang2019review,porte2020wind}. Furthermore, new wind farms often operate under complex and non-ideal conditions~\cite{stevens2017flow,veers2019grand}, including thermal stratification due to diurnal cycles, heterogeneous topography of the land surface, atmosphere-ocean interactions in offshore farms, and farm-to-farm interactions~\cite{porte2020wind}. These factors can influence the characteristics of far-wake turbulence in non-trivial ways. Crucially, existing wake models~\cite{larsen2008wake,bastine2015towards,thogersen2017statistical,mao2018far,gupta2019low,shapiro2017model,gebraad2016wind} are not designed to capture such turbulence, leading to inaccurate predictions, financial losses, and energy security risks~\cite{veers2019grand,stevens2017flow,vermeer2003wind}. Therefore, it is crucial to characterize and understand the far-wake turbulence and to develop predictive models that are applicable under various conditions \cite{han2018atmospheric, porte2020wind, stevens2017flow}.

Recent experiments \cite{heisel2018spectral, coudou2018experimental} and high-fidelity numerical simulations \cite{foti2018similarity,feng2022componentwise} have shown that turbulence generation in the far wakes of wind turbines predominantly results from convective instabilities. Such instabilities, which arise from the shear between the ambient and wake flow, can selectively amplify any upstream velocity fluctuations. As a result there is a broad peak in the turbulent kinetic energy (TKE) spectra at $0.2 \lesssim St \lesssim 0.4$ in the far-wake, where $St\equiv fD/U$ is the Strouhal number, $f$ is the frequency, $D$ is the turbine diameter, and $U$ is the mean incoming velocity at the hub height. However, existing dynamic wake models either neglect this convective instability mechanism \cite{larsen2008wake, gebraad2016wind, gebraad2014control, shapiro2017model, thogersen2017statistical, braunbehrens2019statistical} or can only incorporate it to predict turbulence under simplified inflow conditions~\cite{mao2018far, gupta2019low}. 

Reduced-order models based on modal decomposition, such as proper orthogonal decomposition~\cite{debnath2017towards, bastine2015towards, bastine2018stochastic} and dynamic mode decomposition \cite{iungo2015data, debnath2017towards}, offer a linear, reduced-dimension approximation of the wake flow. These models reconstruct the wake flow field using the coherent structures extracted from time-varying velocity data. Recently, machine-leaning approaches are used in aid of proper orthogonal decomposition to enhance the model accuracy under various inflow and operational conditions~\cite{iungo2022machine,ali2021clustering}. However, such data-driven models either suffer from limited predictive capability~\cite{debnath2017towards, bastine2015towards, bastine2018stochastic} or rely on ad-hoc parameters~\cite{iungo2015data, debnath2017towards, moon2017toward,iungo2022machine}.

Resolvent analysis~\cite{trefethen1993hydrodynamic, jovanovic2005componentwise, schmid2002stability, mckeon2010critical, farrell1993stochastic} is a valuable tool for studying the generation and sustenance of energetic structures in a wide range of fluid systems \cite{taira2017modal}, including wall-bounded flows \cite{hwang2010linear, hwang2010amplification, sharma2013coherent, mckeon2010critical}, turbulent jets \cite{jeun2016input, schmidt2018spectral, garnaud2013preferred}, and wakes \cite{jin2021energy, symon2018non, de2022stability}. This approach, which leverages control theory and is based on the linearized flow equations, involves formulating a transfer function, known as the resolvent operator, between external forcing and the response of a system in the frequency domain~\cite{trefethen1993hydrodynamic, jovanovic2005componentwise, schmid2002stability}. Although the resolvent operator itself is linear, the effect of the nonlinear term is not neglected in the analysis. Its effect is usually modelled via a stochastic forcing term~\cite{hwang2010linear, mckeon2010critical}. Finally, singular value decomposition of the resolvent operator is performed to extract a set of optimal forcing and response modes ranked by their gain. This approach provides the highest-gain modes as the basis for reduced-order models \cite{gomez2016reduced, jovanovic2021bypass, moarref2013model, pickering2021resolvent}.

When the optimal gain significantly exceeds the suboptimal values, referred to as low-rank behavior, the optimal modes closely resemble the energetic structures extracted from data \cite{pickering2021optimal, towne2018spectral}. However, in turbulent flows, where the response is often not low-rank, the presence of a non-trivial structure in the nonlinear forcing terms can lead to discrepancies between the resolvent and data-educed modes \cite{pickering2021optimal, schmidt2018spectral}. Therefore, accurately modeling the nonlinear forcing is necessary to ensure quantitative accuracy in resolvent analysis \cite{mckeon2017engine, towne2018spectral, schmidt2018spectral}. To address this, three types of approaches have been developed: (i) empirical functions \cite{pickering2021resolvent, towne2017statistical}, (ii) data-informed algorithms \cite{zare2017colour, towne2020resolvent, moarref2012model, yim2019self} that utilize partial statistics of the response, and (iii) eddy-viscosity models \cite{morra2019relevance, pickering2021optimal} to account for the dissipative effect of small-scale turbulence~\cite{reynolds1972mechanics, reynolds1967stability}. Recent resolvent analyses of turbulent channel flow \cite{morra2019relevance} and turbulent jets \cite{pickering2021optimal} have demonstrated that incorporating eddy-viscossity models in the resolvent operator can markedly improve the agreement between the modes from the resolvent operator and those extracted via spectral proper orthogonal decomposition (SPOD) \cite{towne2018spectral}. SPOD is a post-processing method to educe the most energetic structures from time-resolved data, while resolvent analysis is a predictive method to obtain most amplified flow structures using only the mean flow data. SPOD modes obtained from LES data, therefore, can be used to assess the accuracy of the predictions from resolvent analysis.

To date, the use of resolvent analysis to study wind-turbine wakes has only been explored by \citet{de2022stability}. They investigated a wind turbine operating within an atmospheric boundary layer and analyzed the three-dimensional wake flow, with a particular focus on the near-wake region. The resolvent operator was formulated under the approximation of streamwise homogeneity, incorporating the mean flow from various cross-sections along the streamwise direction. The resulting resolvent modes exhibited reasonable agreement with the modes extracted from wake flow data using proper orthogonal decomposition, particularly near the rotor. However, the agreement deteriorated in the far-wake region. 

The aim of the present study is to exploit the predictive capabilities of resolvent analysis in the far-wake flow, which predominantly governs the performance and loads of wind farms. In contrast to the approach taken by~\citet{de2022stability}, we consider the wake flow variations in the streamwise direction and approximate to axisymmetric mean flow, which is common in wake models~\cite{jensen1983note,bastankhah2014new,larsen2007dynamic,jonkman2017development}. The primary motivation for this work stems from the utilization of resolvent analysis to develop models capable of predicting turbulence resulting from convective instabilities, such as shown for turbulent jets~\cite{pickering2021resolvent}. It is observed that while the dynamic wake models~\cite{larsen2007dynamic,jonkman2017development} that are widely used in wind farm calculations can predict the mean wake flow and low-frequency wake meandering (i.e., large-scale oscillation induced by atmospheric turbulence) in wind-turbine wakes, their capability in predicting TKE is usually limited~\cite{shaler2021fast}. Reduced-order models based on resolvent analysis of the mean flow can potentially be integrated into dynamic wake models to enhance their applicability. Moreover, resolvent analysis unveils the spectral and spatial characteristics of the upstream fluctuations that have maximum impact on the far-wake dynamics, which could be exploited to develop dynamic control of wind farms.

The rest of this paper is organized as follows. In Sec.~\ref{sec-wakeRes-methods}, we briefly explain the large-eddy simulation (LES), resolvent analysis, and SPOD methods used in this study. In Sec.~\ref{sec-wakeRes-results}, we present the results obtained from resolvent analysis. Specifically, in Sec.~\ref{sec-wakeRes-spec-comp}, we analyze the effect of different velocity component fluctuations in the upstream flow on the TKE generation in the far-wake region. In Sec.~\ref{sec-wakeRes-FR}, we show the optimal inflow perturbations (forcing mode) and the resulting energetic structures in the wake region (response mode). In Sec.~\ref{sec-wakeRes-SPOD-RA}, we compare the modes obtained from resolvent analysis with those extracted from SPOD. In Sec.~\ref{sec-wakeRes-viscEffect}, we analyze the role of eddy-viscosity models in the resolvent analysis of wind turbine wakes.
Finally, in Sec.~\ref{sec-wakeRes-conc}, we present the conclusions and discuss the implications for wind turbine performance.

\section{Methods}\label{sec-wakeRes-methods}

\subsection{Large-eddy simulation}\label{sec-wakeRes-method-LES}

We conduct simulations of a wind turbine operating in an atmospheric boundary layer using SOWFA (simulator for wind farm applications)~\cite{churchfield2012numerical}, which is an LES solver based on the open-source computational fluid dynamics software OpenFOAM 2.4.x~\cite{jasak1996error}. The simulations are performed in two stages: precursor and turbine simulations. In the precursor stage, we simulate a neutral atmospheric boundary layer over a terrain that is representative of level grass plains. The computational domain is a straight channel with dimensions of $2\pi H \times \pi H \times H$ (where $H = 1000$ m) in the streamwise, spanwise and vertical directions, respectively. The domain is discretized using $804 \times 402 \times128$ grid points. The effects of sub-grid scales are modeled using the Lagrangian-averaged scale-dependent dynamic model~\cite{bou2005scale}. For the bottom boundary, a wall model is employed, which accounts for the total viscous and sub-grid scale stresses based on the local-similarity-based logarithmic law~\cite{moeng1984large, kawai2012wall}. The effective roughness height is $0.01$~m to represent the level grass plains. Further numerical details, solver validation, the velocity profiles can be found in~\cite{feng2022componentwise,wang2023implications}. Once the precursor simulation reaches a statistically steady state, we save the velocity fields at the inflow boundary to provide the incoming turbulent flow for the next simulation stage.

In the second stage, we simulate a single wind turbine operating in the fully developed atmospheric boundary layer considered in the precursor simulation. The turbine geometry is based on the NREL 5MW reference turbine~\cite{jonkman2009definition}, which is a three-bladed horizontal-axis turbine with a rotor diameter of $D = 126$ m and a hub height of $h_{hub} = 90$ m~\cite{jonkman2009definition}. The turbine is positioned 10$D$ downstream from the inflow boundary. The turbine is modeled using an actuator disk method with rotation~\cite{wu2011large} and the effects of the nacelle and tower are neglected. This method has shown good agreement with wind tunnel measurements and other high-fidelity numerical simulations in the far-wake region~\cite{feng2022componentwise,porte2011large,stevens2018comparison}. The turbine operates at a tip-speed ratio of 7.55, which corresponds to the peak power coefficient~\cite{jonkman2009definition}. Once the turbine simulation has converged, we collect 2000 snapshots of the instantaneous velocity field with a time step of 2~$s$. The data-collection domain covers $-10 \leq x/D \leq 30$, $-3 \leq y/D \leq 3$, and $-0.7 \leq z/D \leq 2$, where $x$, $y$, and $z$ represent the streamwise, spanwise, and vertical coordinates, respectively. The origin is located at the turbine hub center.

We note that our method does not resolve the tip/root vortices and sub-grid scales. The tip/root vortices are important in the near-wake dynamics. In the far-wake region (the focus of this study), such a LES method is found to well capture the wake dynamics, such as the mean wake profile~\cite{wu2011large} and dominant frequencies of TKE generation~\cite{heisel2018spectral, coudou2018experimental,feng2022componentwise,gupta2019low}. Not including the sub-grid scales lead to a high-frequency cut-off in the TKE spectra, and can also affect the turbulence mixing and hence the mean wake recovery. However, the dominant frequency range of the far-wake TKE generation is significantly lower than the cut-off frequency due to the unresolved scales. Therefore, we consider the influence of these unresolved flow structures on the resolvent analysis not to be significant.

\subsection{Resolvent analysis}\label{sec-wakeRes-method-RA}

Through the use of the Reynolds decomposition to separate the flow variable $\boldsymbol{q}=[\boldsymbol{u}^T,p]^T$, where $\boldsymbol{u}=[u_x,u_y,u_z]^T$ represents the velocity vector and $p$ is the density-normalized pressure, into its mean ($\boldsymbol{Q}$) and fluctuating ($\boldsymbol{q}'$) components, the incompressible Navier--Stokes equations for the fluctuation variables~\cite{mckeon2010critical,hwang2010linear} can be expressed as:
\begin{equation}
	\begin{aligned}
		\boldsymbol{\nabla} \cdot \boldsymbol{u'} &= 0,
	\end{aligned}
	\label{eq_mass}
\end{equation}
\begin{equation}
\begin{aligned}
\frac{\partial \boldsymbol{u}'}{\partial t} + (\boldsymbol{U} \cdot \boldsymbol{\nabla})\boldsymbol{u}' + (\boldsymbol{u}' \cdot \boldsymbol{\nabla})\boldsymbol{U} &= - \boldsymbol{\nabla} p' + \boldsymbol{\nabla} \cdot [\nu_{\rm eff}(\boldsymbol{\nabla} \boldsymbol{u}'+ \boldsymbol{\nabla} \boldsymbol{u}'^T)] + \boldsymbol{f},
\end{aligned}
\label{eq_momentum}
\end{equation}
where $\nu_{\rm eff}=\nu+\nu_t$ is the effective viscosity, $\nu=10^{-5}$~$m^2/s$ is the molecular viscosity, $\nu_t$ is the eddy viscosity and $\boldsymbol{f}=[f_x,f_y,f_z]^T$ represents the external forcing. The eddy-viscosity term mainly accounts for the dissipative effect of the background small-scale turbulence on the large-scale energetic structures, which has been demonstrated to be important in turbulent channel flow~\cite{hwang2010linear,morra2019relevance}, jets~\cite{pickering2021optimal}, and wakes~\cite{de2022stability}. We note that although Eq.~(\ref{eq_momentum}) is linear, the effect of the nonlinear term is approximated by combining the forcing and eddy viscosity terms. Eqs.~(\ref{eq_mass}) and (\ref{eq_momentum}) can be compactly written as
\begin{equation}
	\mathcal{M} \frac{\partial \boldsymbol{q}'}{\partial t} + \mathcal{A}\boldsymbol{q}' =\mathcal{B} \boldsymbol{f},
	\label{eq_compactForm}
\end{equation}
with
\begin{equation}
	\mathcal{M}= \begin{bmatrix} \mathcal{I} & 0 \\ 0 & 0 \end{bmatrix},~
	\mathcal{A}= \begin{bmatrix} \mathcal{L} & -\boldsymbol{\nabla}() \\ \boldsymbol{\nabla} \cdot() & 0 \end{bmatrix} 
	~{\rm and}~
	\mathcal{B}= \begin{bmatrix} \mathcal{I} \\ 0 \end{bmatrix},
\end{equation}
where $\mathcal{I}$ represents the identity operator, $\mathcal{L}=\boldsymbol{U} \cdot \boldsymbol{\nabla}() + ()\cdot \boldsymbol{\nabla}\boldsymbol{U} - \boldsymbol{\nabla} \cdot [\nu_{\rm eff}(\boldsymbol{\nabla}()+ \boldsymbol{\nabla} ()^T)]$.

In general, resolvent analysis is applicable to three-dimensional turbulent flows. In the present study, however, we limit the analysis to the spanwise plane passing through the turbine center and and the mean flow is approximated as axisymmetric in the formulation of the resolvent operator. Thus, the mean flow in our study is only considered in the spanwise plane $z/D = 0$, and the axisymmetric flow approximation means that the streamwise, spanwise and wall-normal velocity components become the streamwise ($u_x$), radial ($u_r$) and azimuthal ($u_\theta$) velocity components, respectively. There are two reasons for this simplification. First, the full three-dimensional flow includes interactions between the wake and the wall turbulence (see Appendix~\ref{apdx_SPODXZ}), which is beyond the scope of the present study. Second, the computational cost of three-dimensional resolvent analysis is markedly higher, requiring the use of either randomized singular value decomposition~\cite{Ribeiro2020Randomized} or variational methods~\cite{Barthel2022Variational}, which introduce numerical approximations.

We note that alternatively the analysis of truly axisymmetric wind turbine wake could be performed for validating the resolvent analysis. However, such an analysis has already been performed by~\citet{mao2018far}. They observed that the response modes to the optimal forcing, which is similar to the resolvent modes calculated in the present study, match well with the response modes to synthetic turbulence. Additionally, the resolvent analysis methodology is well-tested for several ideal flows, including axisymmetric turbulent jets~\cite{pickering2021optimal}. We, therefore, focus on applicability of resolvent analysis for calculation of TKE generation in more practical situation than axisymmetric wakes. We also note that the axisymmetric approximation does not mean that our analysis is only valid for axisymmetric mean flows, such as with uniform inflow boundary conditions. The mean flow on which resolvent analysis is based and the data-driven modes with which resolvent modes will be compared in Sec.~\ref{sec-wakeRes-SPOD-RA} are both obtained from three-dimensional numerical simulations. The present study is thus consistent with existing engineering wake models in which the axisymmetric approximation is made despite being applied to three-dimensional wake flows~\cite{larsen2007dynamic,jonkman2017development}. Additionally, we also ignore the mean azimuthal velocity component, which is known to have only a small effect~\cite{gupta2019low}. In what follows, we use ($U_x$,~$U_r$,~$U_\theta$) for the mean velocity components and ($u_x'$,~$u_r'$,~$u_\theta'$) for the fluctuating velocity components. 

\begin{figure}[htbp]
	\centerline{\includegraphics[width=1\textwidth]{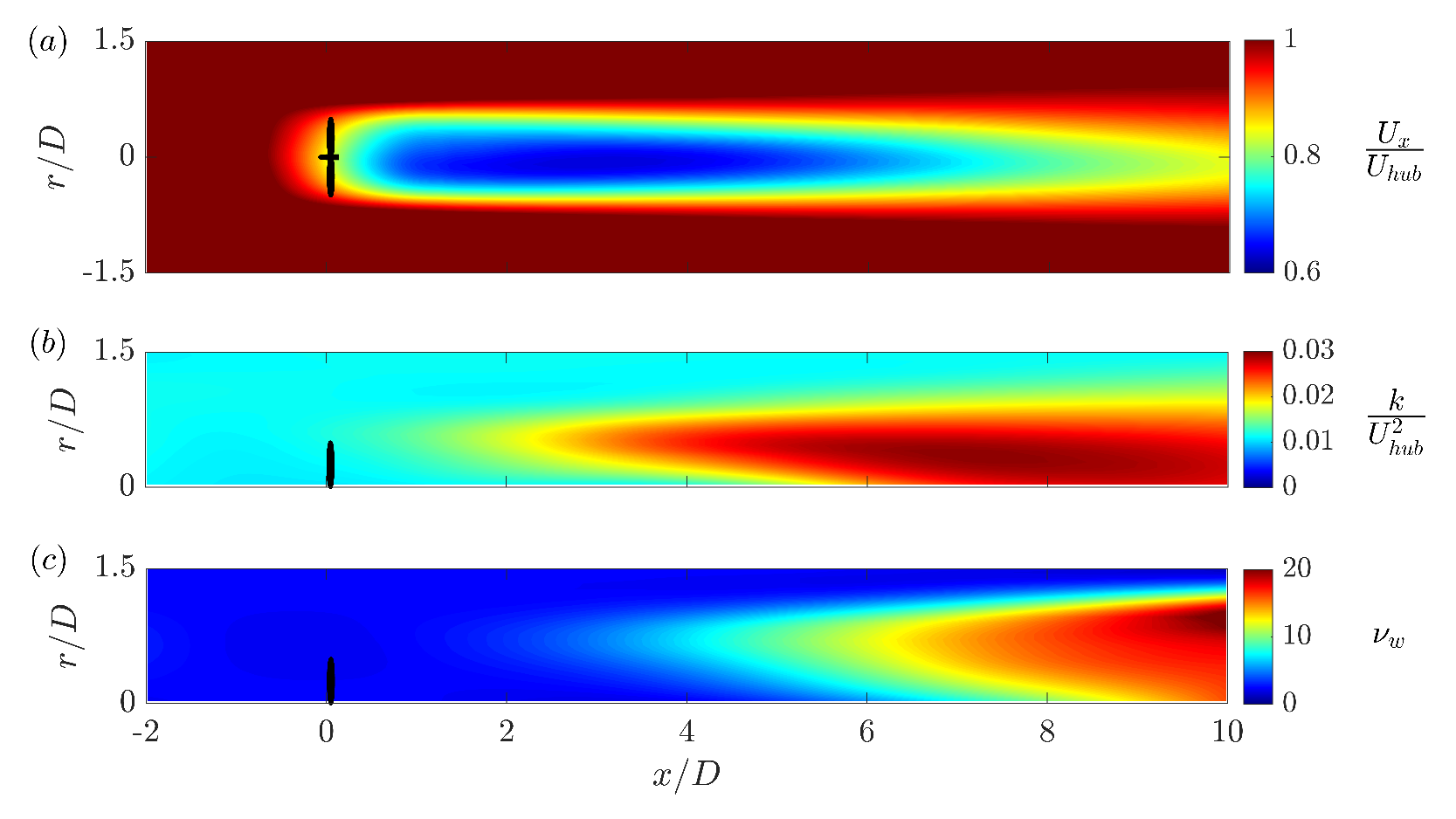}}
	\caption{The LES solutions of ($a$) mean streamwise velocity, ($b$) mean TKE and ($c$) eddy-viscosity field $\nu_w$ (unit: $m^2/s$) in a spanwise plane passing through the turbine center. The locally negative values in ($c$) are set to zero and then smoothed to avoid unphysical gradients. For the smoothing, we use a Savitzky-Golay filter that fits the one-dimensional data in the streamwise and radial directions to a quadratic polynomial using a 7-element sliding window. In these contours as well as the following ones, the black bars indicate the location of turbine.}
	\label{fig_wakeRes_contour_LES}
\end{figure}

Figure~\ref{fig_wakeRes_contour_LES}($a$) shows the mean streamwise velocity, $U_x$, in the spanwise plane $z/D = 0$ of the LES flow field. As the mean flow is nearly axisymmetric, we will perform the following analysis only in the half plane ($r/D\geq0$). It can be seen in Fig.~\ref{fig_wakeRes_contour_LES}($b$) that the mean TKE, $k=\overline{u_x'^2+u_r'^2+u_{\theta}'^2}$, where the overbar denotes time-averaging, in the wake is primarily concentrated in the region $r/D<1$. Based on this, the radial domain size $0 \leq r/D \leq 3$ is chosen for the analysis and prediction of turbulence generation in the far wake. To obtain the resolvent operator, we consider three different models for the effective viscosity $\nu_{\rm eff}$. Specifically, we consider: (i) $\nu_{\rm eff}=\nu$, where $\nu_t=0$, (ii) $\nu_{\rm eff}=\nu+\nu_w$, where $\nu_w$ (see Fig.~\ref{fig_wakeRes_contour_LES}($c$)) is computed as the ratio of the Reynolds shear stress $\overline{u'_x u'_r}$ to the mean shear $d U_x/d r$, and (iii) $\nu_{\rm eff}=\nu+\nu_b$, where $\nu_b=2.52$ $m^2/s$ is calculated as the plane-average value of the ratio of $\overline{u'_x u'_r}$ to $d U_x/d r$ at the hub height in the incoming atmospheric boundary layer. Because the atmospheric boundary layer is considered to be horizontally homogenous, $\nu_b$ calculated from the mean flow is a constant. The influence of $\nu_w$ and $\nu_b$ will be assessed in Sec.~\ref{sec-wakeRes-viscEffect}.

For such a statistically steady, axisymmetric flow, the fluctuating variables $\boldsymbol{q'}=[u_x',u_r',u_{\theta}',p']^T$ and the external forcing $\boldsymbol{f}=[f_x,f_r,f_\theta]^T$ can be decomposed using the Fourier basis
\begin{equation}
	\begin{aligned}
		\boldsymbol{q}'(x,r,\theta,t) &= \sum_m\sum_\omega \hat{\boldsymbol{q}}_{m\omega}(x,r)e^{i(m\theta-\omega t)},
	\end{aligned}
	\label{}
\end{equation}
\begin{equation}
	\begin{aligned}
		\boldsymbol{f}(x,r,\theta,t) &= \sum_m\sum_\omega \hat{\boldsymbol{f}}_{m\omega}(x,r)e^{i(m\theta-\omega t)},
	\end{aligned}
\label{}
\end{equation}
where $m$ is the azimuthal wavenumber (integer) and $\omega$ is the angular frequency. By rearranging Eq.~(\ref{eq_compactForm}), we obtain an input-output equation~\cite{jovanovic2005componentwise,hariharan2021well}
\begin{equation}
	\begin{aligned}
		(-i\omega\mathcal{M} + \mathcal{A}_m) \hat{\boldsymbol{q}}_{m\omega} &= \mathcal{B} \hat{\boldsymbol{f}}_{m\omega},
	\end{aligned}
	\label{}
\end{equation}
\begin{equation}
\begin{aligned}
\hat{\boldsymbol{y}}_{m\omega} &= \mathcal{C} \hat{\boldsymbol{q}}_{m\omega},
\end{aligned}
\label{}
\end{equation}
where $\hat{\boldsymbol{f}}_{m\omega}$ and $\hat{\boldsymbol{y}}_{m\omega}$ represent the input and output, respectively, $\mathcal{C}= \begin{bmatrix} \mathcal{I} & 0 \end{bmatrix}$, and $\mathcal{A}_m$ is the frequency-independent linearized Navier-Stokes operator. Dirichlet boundary conditions $\boldsymbol{u}'=p'=0$ are imposed at the inflow boundary~\cite{garnaud2013preferred,jin2020feedback} and the lateral boundary~\cite{batchelor1962analysis,oberleithner2014mean}. At the outflow boundary, a mixed boundary condition that combines the velocity and pressure is imposed~\cite{garnaud2013preferred,jin2020feedback}
\begin{equation}
	\frac{1}{Re}\frac{\partial \boldsymbol{u}'}{\partial n} - p'\boldsymbol{n} = 0.
	\label{}
\end{equation}
The boundary condition along the turbine centerline is determined based on $m$: $\partial u'_x/\partial r = u'_r = u'_\theta = \partial p'/\partial r = 0$ for even $m$ and $ u'_x = \partial u'_r/\partial r = \partial u'_\theta/\partial r = \partial p'/\partial r = 0$ for odd $m$~\cite{matsushima1995spectral,garnaud2013preferred}. These boundary conditions are incorporated into the matrices $\mathcal{M}$, $\mathcal{A}_m$, and $\mathcal{B}$. To discretize the resolvent operator, we use a second-order central difference scheme on uniform grids.

The relationship between the input and output can also be expressed as $\hat{\boldsymbol{y}}_{m\omega} = \mathcal{H}_{m\omega} \hat{\boldsymbol{f}}_{m\omega}$,
where $\mathcal{H}_{m\omega}  = \mathcal{C}(-i\omega\mathcal{M} + \mathcal{A}_m)^{-1} \mathcal{B}$ is the resolvent operator.
The singular value decomposition of $\mathcal{H}_{m\omega}=\mathcal{U} \Sigma \mathcal{V}^*$ provides a set of forcing modes ($\mathcal{V}=[\boldsymbol{v}_1,\boldsymbol{v}_2,\cdots,\boldsymbol{v}_N]$) and response modes ($\mathcal{U}=[\boldsymbol{u}_1,\boldsymbol{u}_2,\cdots,\boldsymbol{u}_N]$) ranked according to the amplitude gain ($\Sigma={\rm diag}(\sigma_1,\sigma_2,\cdots,\sigma_N)$) between them.
In the present study, resolvent analysis is performed over the range $0.05 \leq St \leq 1$, where $St=\omega D/2\pi U_{\text{hub}}$. This $St$ range can cover the dominant frequencies of far-wake TKE generation~\cite{mao2018far,foti2018similarity,feng2022componentwise}.

\begin{figure}[htbp]
	\centerline{\includegraphics[width=1\textwidth]{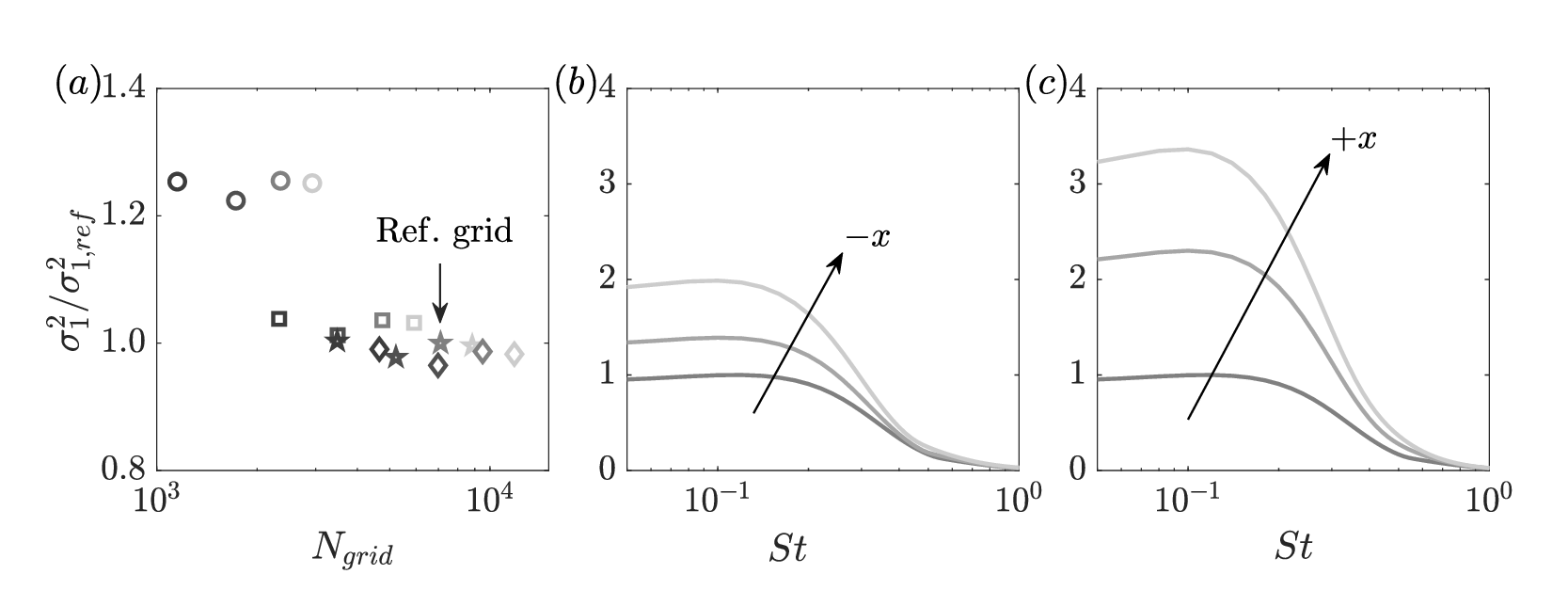}}
	\caption{The optimal resolvent gains ($\sigma_1$) calculated using the $\nu_b$-model: ($a$) The peak value of $\sigma_1$ over all frequencies for different combinations of streamwise ($N_x$) and spanwise ($N_r$) grid cell counts. The circles, squares, stars and diamonds represent $N_x=48,97,145$ and $194$ respectively, and from the dark to light gray represents $N_r=24,36,49$ and $61$ respectively. ($b$) $\sigma_1$ spectra for different upstream extents of the domain. From the dark to light gray, the data represents $2D$, $5D$ and $10D$, respectively. ($c$) $\sigma_1$ spectra for different downstream extents of the domain. From the dark to light gray, the data represents $10D$, $20D$ and $30D$, respectively. These gains are normalized by the reference value $\sigma_{1,{\rm ref}}$ (pointed out by the arrow in ($a$)), which is calculated using the grid resolution $N_{grid}=N_x \times N_r = 145 \times 49 =7105$ and the streamwise domain size $-2 \leq x/D \leq10$.}
	\label{fig_wakeRes_spectra_sizeEffect}
\end{figure}

We investigate the dependence of resolvent analysis on the grid resolution and domain size in the streamwise direction. These calculations are performed with the $\nu_b$-model. Different combinations of streamwise ($N_x=48, 97, 145$, and $194$) and radial ($N_r=24, 36, 49$, and $61$) grid cell counts are tested for a streamwise domain size of $-2 \leq x/D \leq 10$. Figure~\ref{fig_wakeRes_spectra_sizeEffect}($a$) shows that the resolvent gains approximately converge with a grid resolution of $N_{\text{grid}}=N_x \times N_r = 145 \times 49 = 7105$. There is further small reduction in the resolvent gain with the increasing grid numbers, which is a result of increasing viscous dissipation with grid resolution. This reduction is not significant and thus the grid size (marked as Ref. grid in the figure) is adopted for the subsequent resolvent analysis. Next, the effect of domain size is examined by varying the upstream and downstream extents. Fig.~\ref{fig_wakeRes_spectra_sizeEffect} reveals that increasing either the upstream (Fig.~\ref{fig_wakeRes_spectra_sizeEffect}($b$)) or downstream (Fig.~\ref{fig_wakeRes_spectra_sizeEffect}($c$)) extent leads to higher resolvent gains. This behavior, observed previously in turbulent jets~\cite{garnaud2013preferred}, can be attributed to convective non-normality~\cite{schmidt2018spectral}. The response modes grow as they advect downstream, causing the larger domains to include longer tails of the response modes, particularly for the low-frequency modes which do not dissipate. This leads to a higher apparent gain but has a negligible influence on the shape of the gain spectra (Fig.~\ref{fig_wakeRes_spectra_sizeEffect}($b$) and ($c$)) as well as the resolvent modes (not shown here). Based on these results and the fact that the TKE generation mainly occurs before $x/D=10$, an upstream extent of 2$D$ and a downstream extent of 10$D$ are selected. 

\subsection{Spectral proper orthogonal decomposition}\label{sec-wakeRes-method-SPOD}

SPOD is used to extract energetic modes at various frequencies from the time-resolved simulation data. These data-driven modes provide a basis for evaluating the response modes predicted by resolvent analysis~\cite{towne2018spectral,schmidt2018spectral,pickering2021optimal}. Previous studies by \citet{schmidt2018spectral} and \citet{pickering2021optimal} have used SPOD to analyze three-dimensional axisymmetric turbulent jets in cylindrical coordinates. In a similar spirit, we adopt a comparable approach for the wake flow data in a two-dimensional plane, $-2 \leq x/D \leq 10$ and $-3 \leq r/D \leq 3$. This domain corresponds to the one used in the resolvent analysis, ensuring consistency between the two methods.

In order to separate the modes with odd and even integer values of $m$, we decompose the instantaneous velocity fields into their symmetric and anti-symmetric components:
\begin{equation}
	\boldsymbol{u}_{sym}=\frac{\boldsymbol{u}(r)+\boldsymbol{u}(-r)}{2} 
	~{\rm and}~
	\boldsymbol{u}_{anti}=\frac{\boldsymbol{u}(r)-\boldsymbol{u}(-r)}{2}.
\end{equation}
We assume that the velocity field $(u_{x,\text{anti}}, u_{r,\text{sym}}, u_{\theta,\text{sym}})$ generates modes with odd integer values of $m$. Proper orthogonal decomposition of three-dimensional wake flow shows that the modes with $|m|\geq 3$ have negligible contribution to the TKE. Therefore, the obtained SPOD modes can be directly compared with the resolvent modes at $|m| = 1$. We use 1024 snapshots uniformly-spaced at a time step of $0.2$ s and a Hamming window with a 50$\%$ overlap to obtain the SPOD modes.

\section{Results and discussion}\label{sec-wakeRes-results}
\subsection{Effect of different velocity components and azimuthal wavenumbers on resolvent gains}\label{sec-wakeRes-spec-comp}

\begin{figure}[htbp]
	\centerline{\includegraphics[width=1\textwidth]{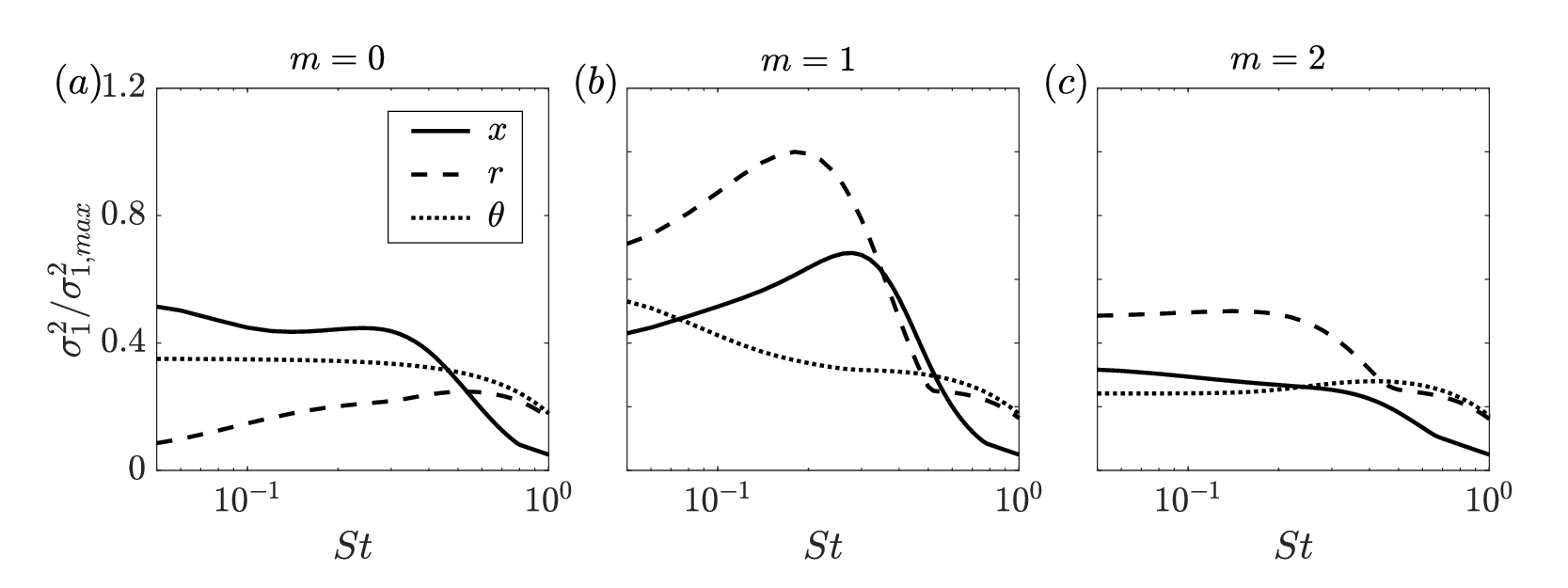}}
	\caption{Spectra of the optimal resolvent gain $\sigma_1$ for different velocity components: $x$, $r$ and $\theta$, and different azimuthal wavenumbers: ($a$) $m=0$ ($b$) $m=1$ and ($c$) $m=2$. The resolvent gains are computed using the $\nu_w$-model, and their spectra are normalized by the maximum value of the $r$ component at $m=1$.}
	\label{fig_wakeRes_spectra_compEffect}
\end{figure}

We investigate the effectiveness of forcing in different directions (i.e., $x$, $r$, and $\theta$) on the TKE generation in the wake flow; this is similar to the approach outlined in~\cite{jovanovic2005componentwise}. Throughout this section, as well as in Secs.~\ref{sec-wakeRes-FR} and \ref{sec-wakeRes-SPOD-RA}, all resolvent analysis calculations are performed using the $\nu_w$-model for the eddy viscosity. In Fig.~\ref{fig_wakeRes_spectra_compEffect}, we present the spectra of the optimal resolvent gain ($\sigma_1$) corresponding to the three forcing components at ($a$) $m=0$, ($b$) $m=1$ and ($c$) $m=2$. Because the wake flow is approximated to be non-rotating, the results for positive and negative $m$ are identical. We find that $\sigma_1$ has the highest values at $m=1$, which is consistent with the observations made in studies on axisymmetric turbulent jets~\cite{schmidt2018spectral} and wind-turbine wakes~\cite{mao2018far,de2022stability}. Specifically,~\citet{mao2018far} found that $m=1$ modes are optimally amplified by the wake shear and thus dominate the far-wake region of wind turbines. These findings indicate that $m=1$ modes are most unstable in the far-wake region, as also found in~\cite{gupta2019low}. Therefore, in this study, we mainly focus on the resolvent modes at $m=1$.

At $m=1$, the $r$ forcing component exhibits the highest $\sigma_1$ value at $St\approx0.2$. This emphasizes the dominant role of the $r$ forcing component, supporting the observation made by~\citet{feng2022componentwise} that, compared to the streamwise component, the cross-stream components of velocity fluctuations induce stronger TKE generation in the far wake. The second-highest resolvent gain is observed in the $x$ forcing component. Velocity fluctuations in this direction are known to primarily contribute to the TKE of the inflow atmospheric boundary layer~\cite{smits2011high,crespo1996turbulence,panofsky1984atmospheric}. Therefore, we consider both the $x$ and $r$ forcing components to be important in far-wake TKE generation. This is consistent with the findings of \citet{mao2018far}, who observed that the optimal inflow perturbations for axisymmetric wind-turbine wake flows are mainly in the $x$- and $r$-directions.

\subsection{Forcing and response modes from resolvent analysis}\label{sec-wakeRes-FR}

\begin{figure}[htbp]
	\centerline{\includegraphics[width=1\textwidth]{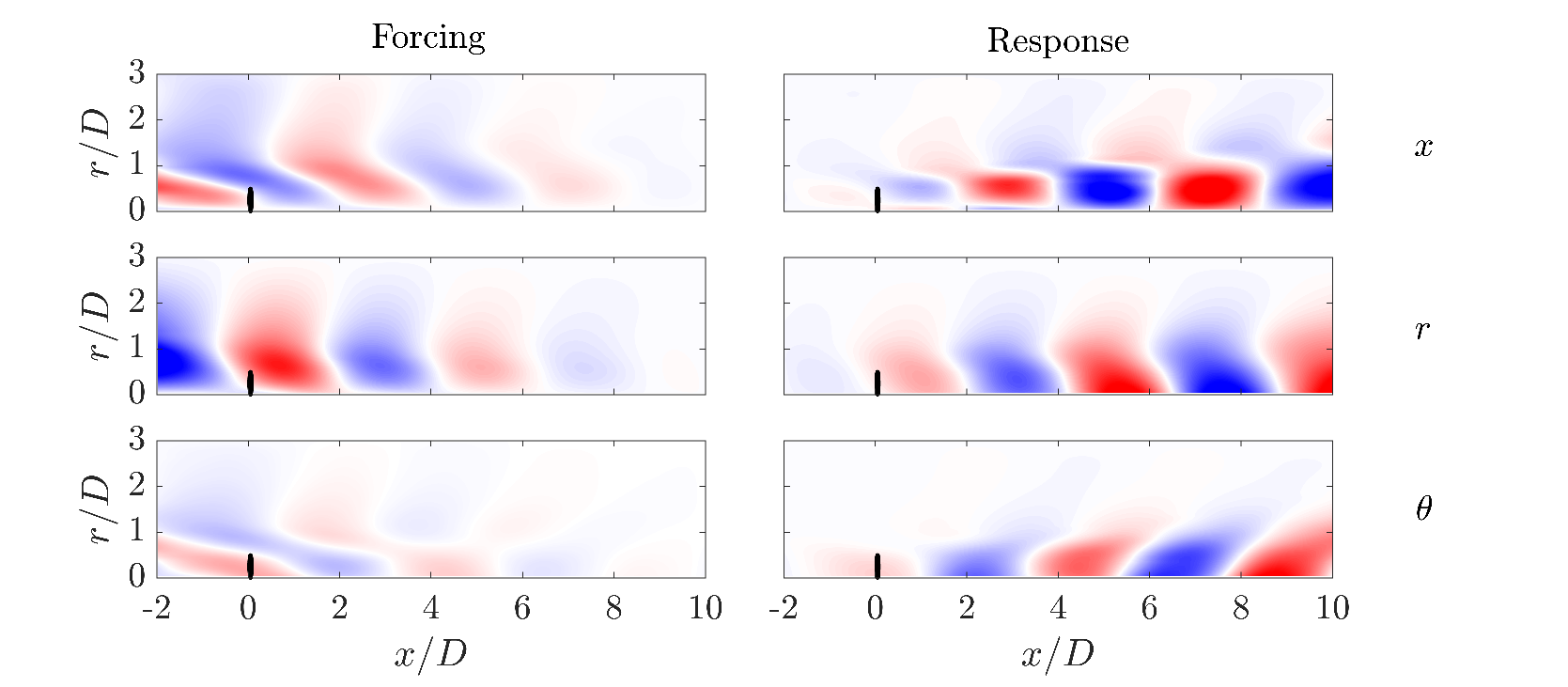}}
	\caption{The optimal forcing (left) and response (right) modes (real part, $m=1$) for different velocity components. The resolvent modes are calculated using the $\nu_w$-model. The first to third rows show the results for the $x$, $r$ and $\theta$ velocity components, respectively. The results are normalized by the maximum magnitude of the modes over all the velocity components, frequencies and space. All contours in this paper are such that the white regions indicate negligible magnitude while the transition from white to red and white to blue indicate changes in values to 0.5 and -0.5, respectively.}
	\label{fig_wakeRes_mode_FR_comp}
\end{figure}

In Fig.~\ref{fig_wakeRes_mode_FR_comp}, we present the three components of the leading forcing and corresponding response modes at $St=0.2$. In all three components, the forcing is observed to be located upstream of the response, indicating the presence of a convective instability mechanism. Furthermore, the $r$-velocity component exhibits the highest amplitudes in the forcing modes. This observation is consistent with the strongest influence of the $r$-forcing component on the far-wake TKE generation, as shown in Fig.~\ref{fig_wakeRes_spectra_compEffect}.

\begin{figure}[htbp]
	\centerline{\includegraphics[width=1\textwidth]{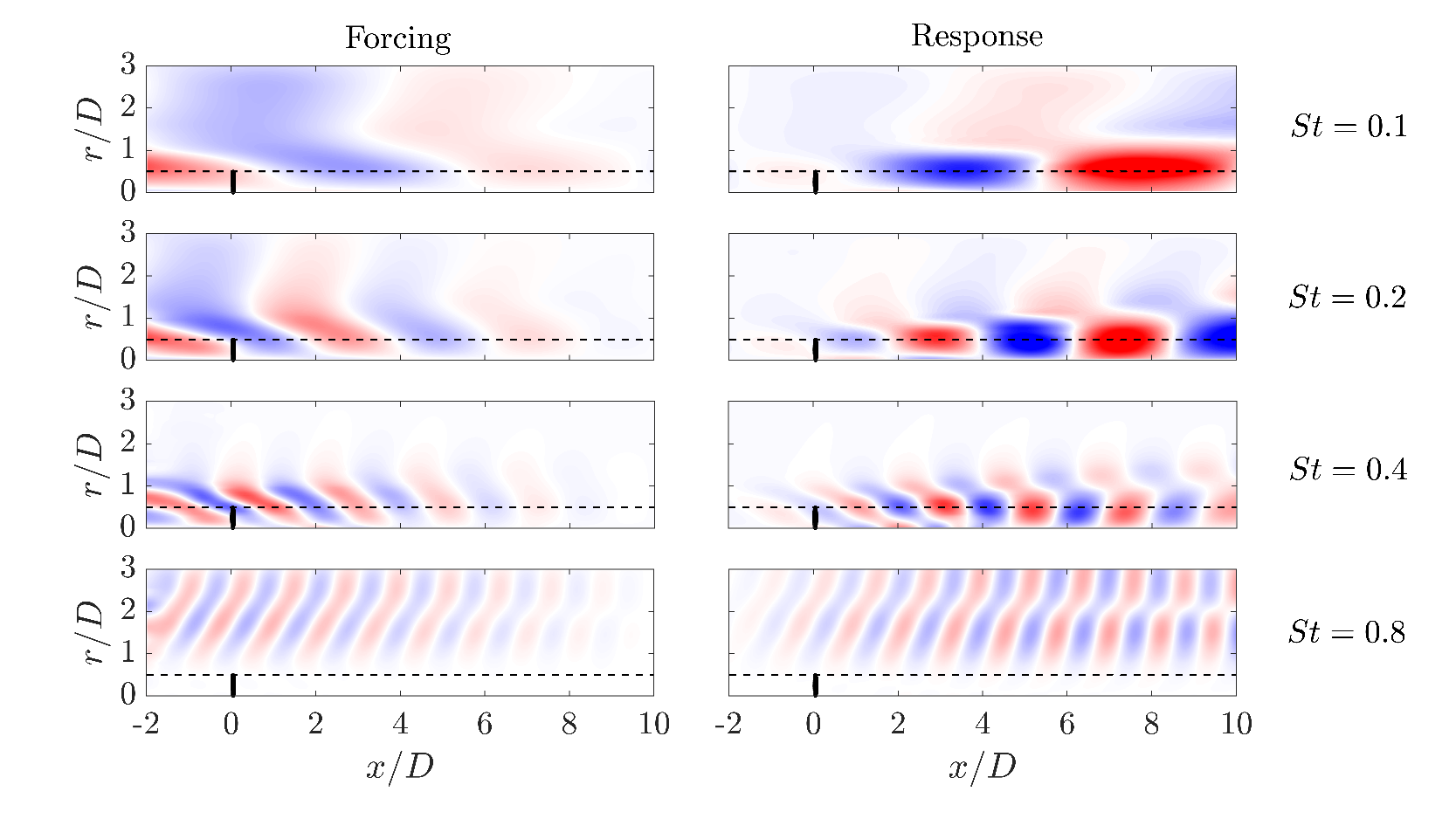}}
	\caption{The optimal forcing (left) and response (right) modes (real part, $m=1$). The resolvent modes are calculated using the $\nu_w$-model. The first to fourth rows present results at $St=0.1$, $0.2$, $0.4$ and $St=0.8$, respectively. The dashed lines represent $r/D=0.5$, which corresponds to the radial location of the high mean-shear region in the wake flow.}
	\label{fig_wakeRes_mode_FR_freq}
\end{figure}

In Fig.~\ref{fig_wakeRes_mode_FR_freq}, we show the influence of frequency on the mode shape by presenting the $x$-velocity component of the leading forcing-response mode pair at $St=0.1$, $0.2$, $0.4$, and $0.8$. We find that the streamwise wavelength of the forcing and response modes decreases as the frequency increases, indicating an inverse proportionality between the wavelength and frequency as expected. The radial distribution of the forcing and response modes varies with frequency. At low frequencies ($St=0.1$, $0.2$, and $0.4$), the forcing modes are in the upstream region (not in the wake shear region), while the response modes are concentrated in the region of high mean shear, around $r/D=0.5$ (indicated by the dashed lines). At $St=0.8$, both the forcing and response modes are concentrated in the free-stream region ($r/D > 0.5$). This distinction highlights two types of  modes. The first type is generated by the convective instability mechanism, such as at $St=0.1$, $0.2$ and $0.4$. The response modes are mainly confined to the turbine wake region with maximum amplitudes at $x/D\approx7$, coinciding with the region of the maximum mean TKE as can be seen in Fig.~\ref{fig_wakeRes_contour_LES}($b$). The second type corresponds to the advection and diffusion of perturbations in the free stream~\cite{jin2021energy} and is exemplified by the mode at $St=0.8$.

We also observe that the forcing modes are tilted in the direction opposite to the mean flow. This is indicative of the Orr mechanism, which manifests as non-modal algebraic amplitude growth~\cite{butler1992three}. Previous resolvent analysis in channel flows~\cite{hwang2010amplification} and jets~\cite{schmidt2018spectral,pickering2020lift} also uncovered the role of the Orr mechanism in amplitude growth of fluctuations. The tilting of the forcing modes seems to be more pronounced at lower frequencies, which agrees with the observations by~\citet{pickering2020lift} that the Orr mechanism is particularly important at low frequencies.~\citet{garnaud2013preferred} concluded for jet flows that non-modal growth via the Orr mechanism can combine with the convective Kelvin-Helmholtz instability mechanism to enhance amplitude growth.In this study, the response modes resemble the flow structures generated via Kelvin-Helmholtz instability. The present results therefore also indicate a combined role of the Orr and convective instability mechanisms for TKE generation in the far-wake of wind turbines.


\subsection{Comparison of SPOD and resolvent analysis}\label{sec-wakeRes-SPOD-RA}

\begin{figure}[htbp]
	\centerline{\includegraphics[width=1\textwidth]{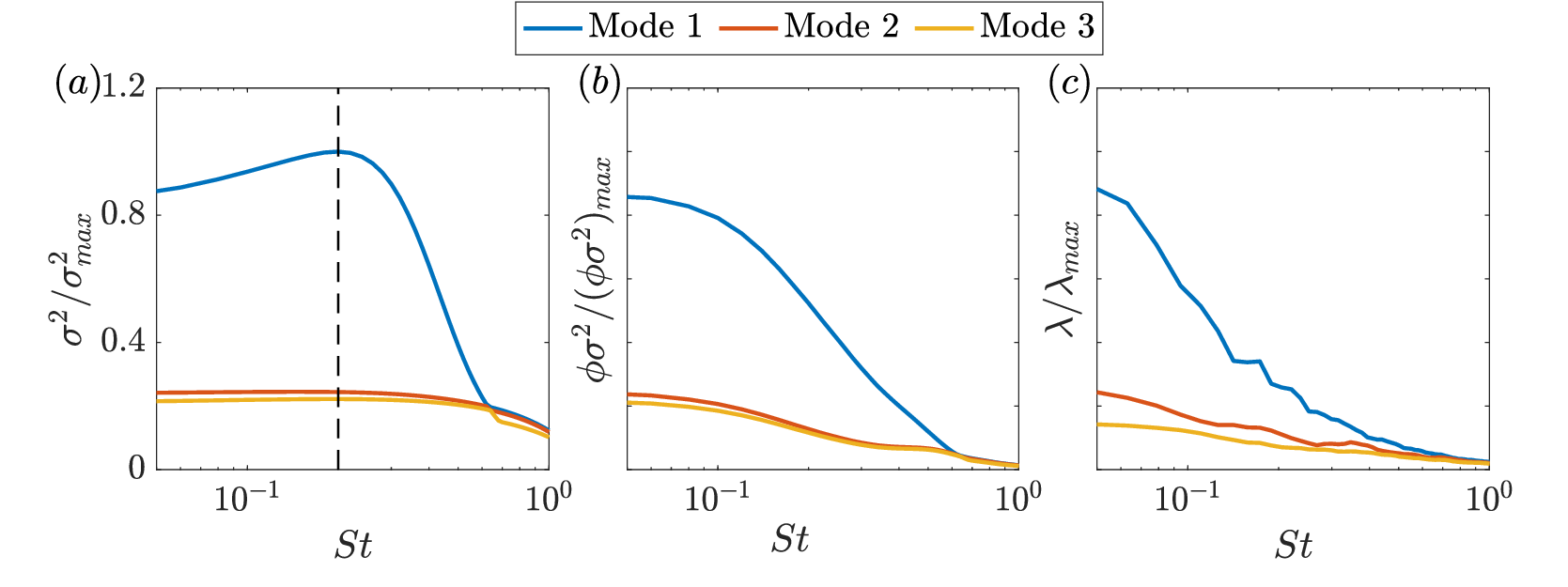}}
	\caption{The spectra of ($a$) resolvent gains (the wake flow), ($b$) the product of TKE (the atmospheric boundary layer inflow) and resolvent gains (the wake flow) and ($c$) SPOD eigenvalues (the wake flow). The resolvent gains are calculated using the $\nu_w$ model. These spectra are normalized by the maximum value in each subfigure. The results for the first three modes at $m=1$ are presented. The dashed line in ($a$) denotes $St=0.2$.}
	\label{fig_wakeRes_spectra_RA-SPOD}
\end{figure}

In order to validate the resolvent analysis, we compare normalized TKE spectra for the dominant modes as predicted by resolvent analysis with those obtained using SPOD. We first show the resolvent gain spectra for the first three modes in Fig.~\ref{fig_wakeRes_spectra_RA-SPOD}($a$). The gain spectra show the frequency-dependent energy amplification of upstream fluctuations. The separation between mode 1 and mode 2 results suggests the dominance of the leading mode over the sub-optimal modes, indicating low-rank behavior in the wake dynamics. The peak frequency, $St\approx0.2$, of the resolvent gains aligns with the dominant frequency range ($0.2 \lesssim St \lesssim 0.4$) associated with far-wake TKE generation~\cite{heisel2018spectral,coudou2018experimental,foti2018similarity,feng2022componentwise}. We approximate the predicted TKE spectra from resolvent analysis, shown in Fig.~\ref{fig_wakeRes_spectra_RA-SPOD}($b$), as the product of the resolvent gain spectra with the inflow TKE spectra, denoted as $\phi=\phi_{u_x}+\phi_{u_r}+\phi_{u_\theta}$, where $\phi_{u_x}$, $\phi_{u_r}$, and $\phi_{u_\theta}$ represent the power spectral densities of the streamwise, radial, and azimuthal velocity fluctuations in the inflow, respectively. Finally, Fig.~\ref{fig_wakeRes_spectra_RA-SPOD}($c$) shows the SPOD eigenvalue spectra (calculated using singular value decomposition of the cross-spectral density tensor [46, 83]) for the first three leading modes (mode 1, 2, and 3). We observe that the normalized TKE spectra obtained from SPOD agrees well with that predicted from the resolvent analysis. These results, therefore, support the use of resolvent analysis for calculating the TKE generation in the far-wake regions of wind turbines.

\begin{figure}[htbp]
	\centerline{\includegraphics[width=1\textwidth]{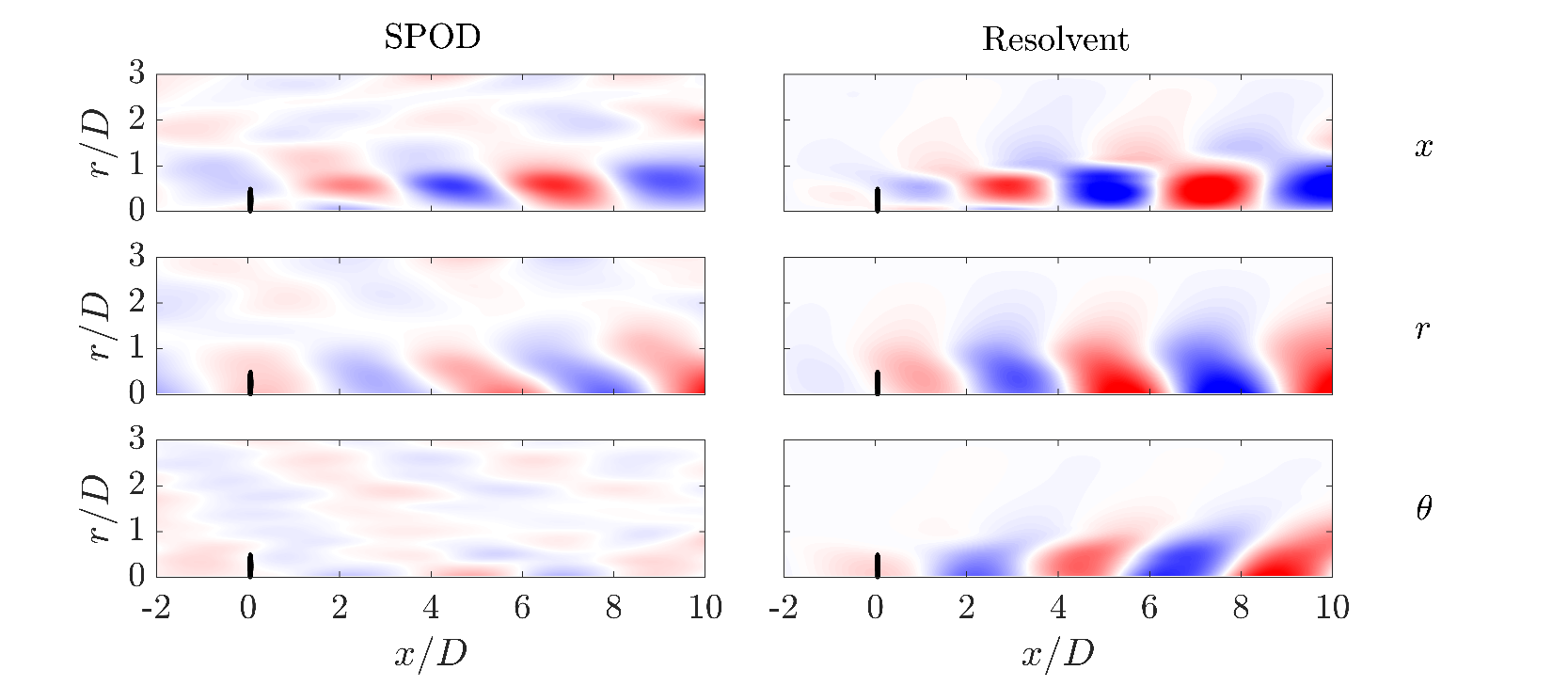}}
	\caption{The optimal SPOD (left) and resolvent response (right) modes (real part, $m=1$) for different velocity components as indicated. The resolvent modes are calculated using the $\nu_w$-model.}
	\label{fig_wakeRes_mode_SPOD-RA_comp}
\end{figure}

In Fig.~\ref{fig_wakeRes_mode_SPOD-RA_comp}, we compare the leading SPOD and resolvent (response) modes at $St=0.2$. For the $x$- and $r$-velocity components, there is a good agreement between the resolvent and SPOD modes. Both modes exhibit similar wavelengths and shapes. However, for the $\theta$-velocity component, there is a discrepancy between the resolvent and SPOD modes. The SPOD mode does not clearly show magnitude peaks in the high mean-shear region ($r/D \approx 0.5$), as observed in the resolvent mode. This mismatch may be attributed to (i) the limitation of the axisymmetry approximation used in this study and (ii) the influence of incoming atmospheric boundary layer fluctuations on the SPOD modes, which is not accounted for in the resolvent modes. Considering the dominant role of the $x$- and $r$-velocity components (see Sec.~\ref{sec-wakeRes-spec-comp}), we assume the mismatch in the $\theta$-velocity component to be of low significance.

\begin{figure}[htbp]
	\centerline{\includegraphics[width=1\textwidth]{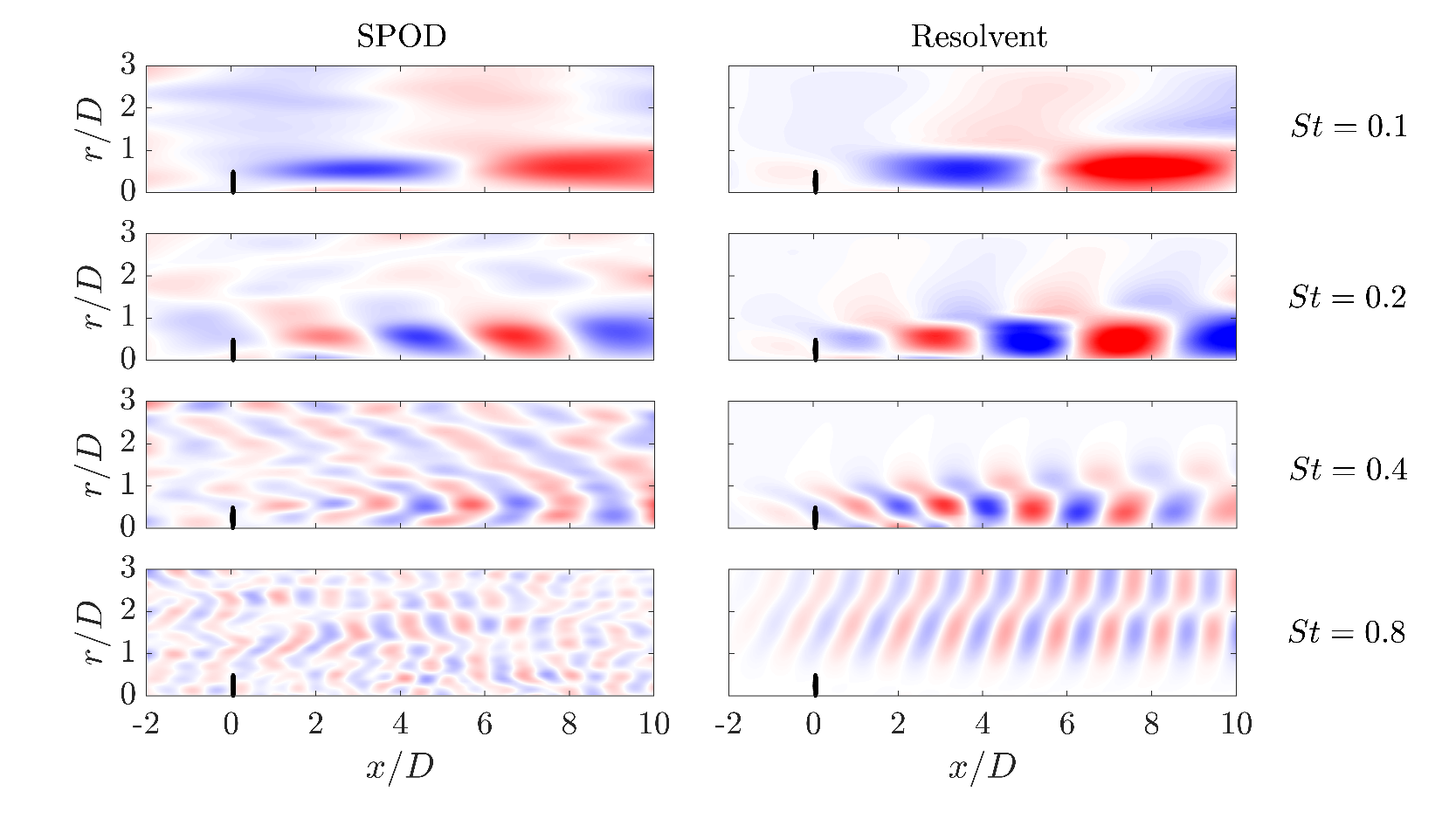}}
	\caption{The optimal SPOD (left) and resolvent response (right) modes (real part, $m=1$) for different $St$ as indicated. The resolvent modes are calculated using the $\nu_w$-model.}
	\label{fig_wakeRes_mode_SPOD-RA_freq}
\end{figure}

To examine the effect of frequency on the mode agreement, we present the $x$-component of the SPOD and resolvent response modes at $St=0.1$, $0.2$, $0.4$, and $0.8$ in Fig.~\ref{fig_wakeRes_mode_SPOD-RA_freq}. At $St=0.1$, $0.2$, and $0.4$, there is a reasonable agreement between the SPOD and resolvent modes. However, at $St=0.8$, the SPOD mode becomes too scattered to be compared with the resolvent modes. As shown in Sec.~\ref{sec-wakeRes-FR}, at $St=0.8$, the far-wake TKE generation is no longer dominated by the convective instability mechanism. Considering the reasonable agreement at frequencies such as $St=0.1$, $0.2$, and $0.4$, which correspond to relatively high resolvent gains, we consider that the resolvent modes effectively capture the energetic structures in the far wake generated through convective instabilities. The mode agreement will be further evaluated in Fig.~\ref{fig_wakeRes_alignment_RA-SPOD}.

\subsection{Effect of the eddy-viscosity model}\label{sec-wakeRes-viscEffect}

\begin{figure}[htbp]
	\centerline{\includegraphics[width=1\textwidth]{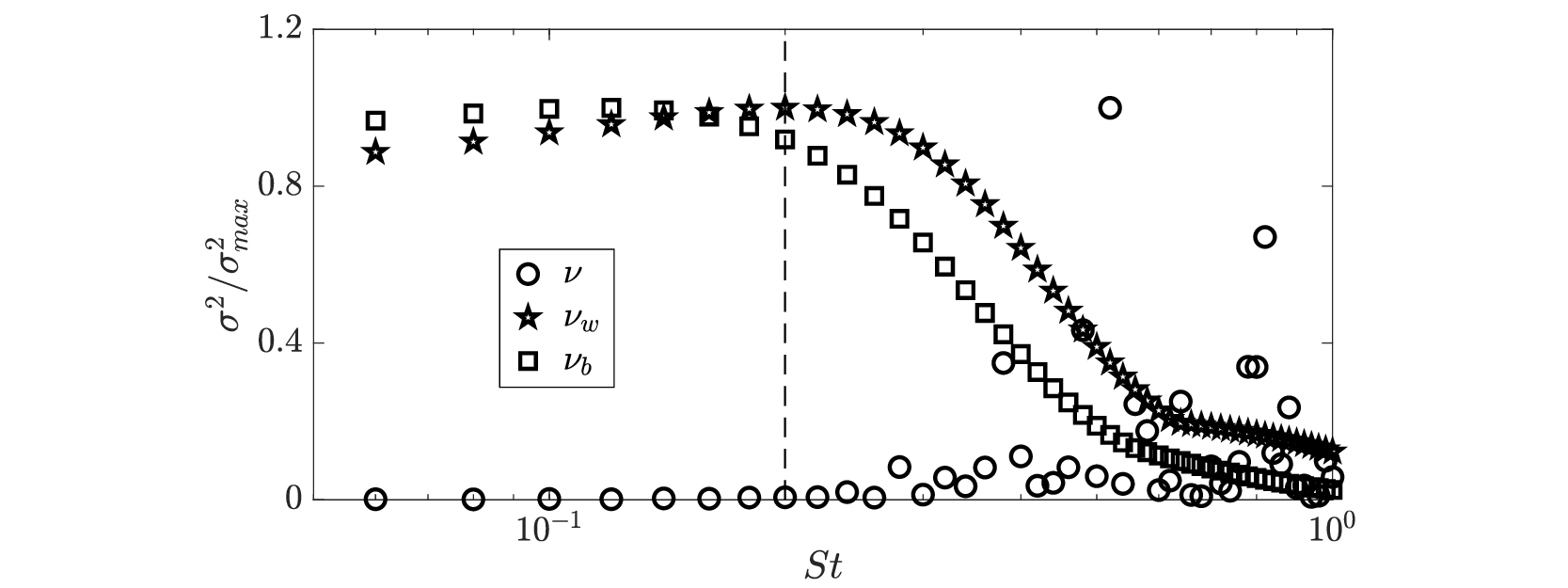}}
	\caption{The spectra of optimal resolvent gains for the $\nu$ (circles), $\nu_w$ (stars) and $\nu_b$ (squares) models for the eddy viscosity. These spectra are normalized by their maximum value. The dashed line denotes $St=0.2$.}
	\label{fig_wakeRes_spectra_viscEffect}
\end{figure}

We investigate the influence of eddy-viscosity models on the performance of resolvent analysis. In Fig.~\ref{fig_wakeRes_spectra_viscEffect}, we compare the spectra of optimal resolvent gains when different models are used in the resolvent analysis. When the eddy viscosity is included (either as $\nu_w$ or $\nu_b$), the resolvent gain spectra exhibits a broad peak centered around $St\approx0.2$. This corresponds to the dominant frequency range ($0.2 \lesssim St \lesssim 0.4$) for the TKE generation as observed in recent studies~\cite{heisel2018spectral,coudou2018experimental,foti2018similarity,feng2022componentwise}. When the eddy visosity is neglected, i.e. $\nu_{\rm eff} = \nu$, the resolvent gain spectra shows unphysical behavior characterized by several sharp peaks at higher frequencies. The inclusion of eddy viscosity effectively captures the influence of background small-scale turbulence on the energetic structures~\cite{pickering2021optimal,hwang2010linear}. Furthermore, compared to the $\nu_b$-model results, the $\nu_w$-model results show a peak frequency that is closer to the expected dominant frequency range, indicating a slight improvement in the spectral characteristics.

\begin{figure}[htbp]
	\centerline{\includegraphics[width=1\textwidth]{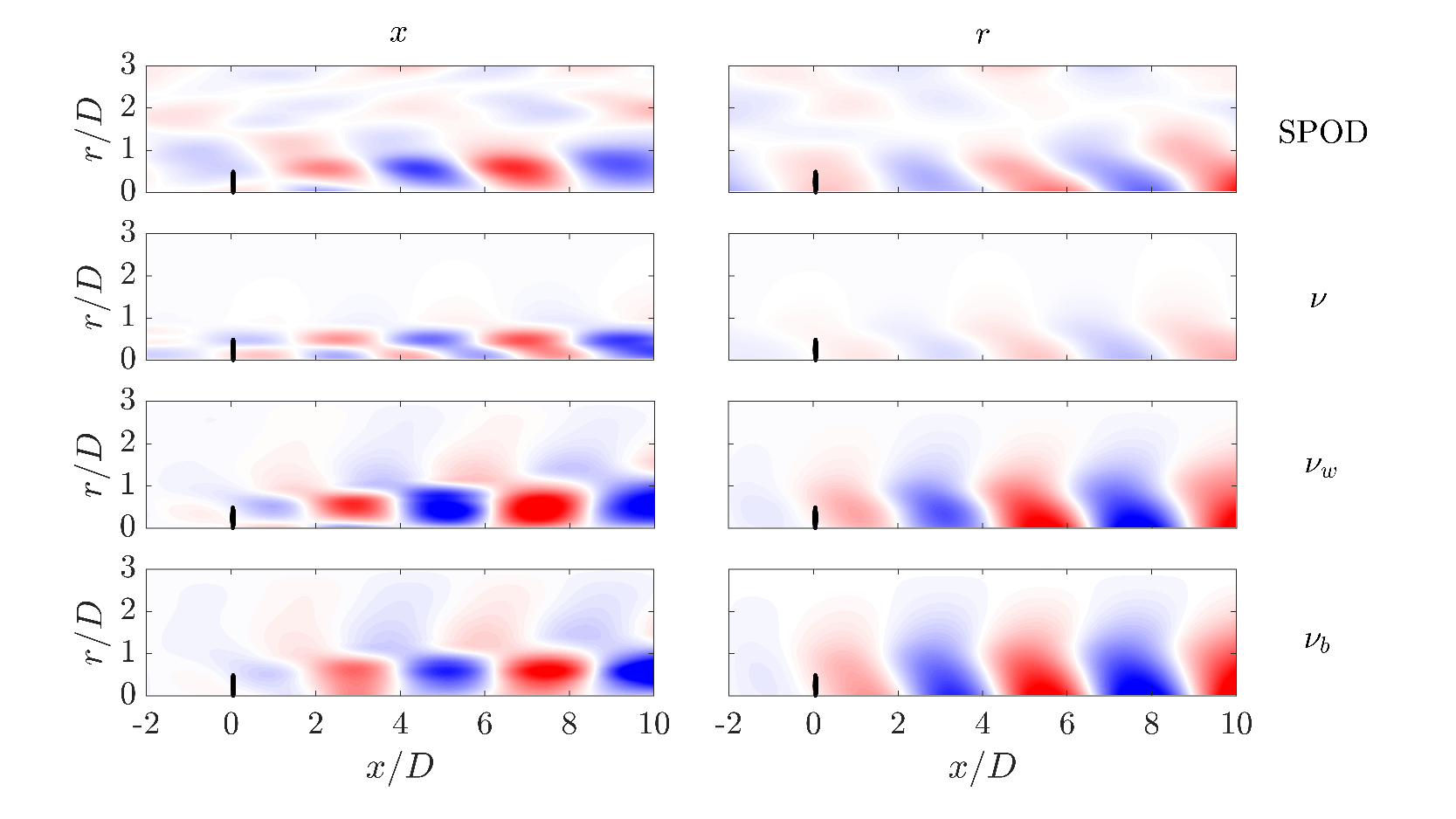}}
	\caption{Dominant SPOD modes (the first row) and the optimal resolvent response modes for three eddy-viscosity models (the second to fourth rows) at $m=1$ . The real part of the streamwise (left) and radial (right) velocity components at $St=0.2$ are presented and normalized by the maximum value over all velocity components at each row. The second to fourth rows present results for the $\nu$, $\nu_w$ and $\nu_b$ models, respectively.}
	\label{fig_wakeRes_mode_SPOD-RA_viscEffect}
\end{figure}

\begin{figure}[htbp]
	\centerline{\includegraphics[width=1\textwidth]{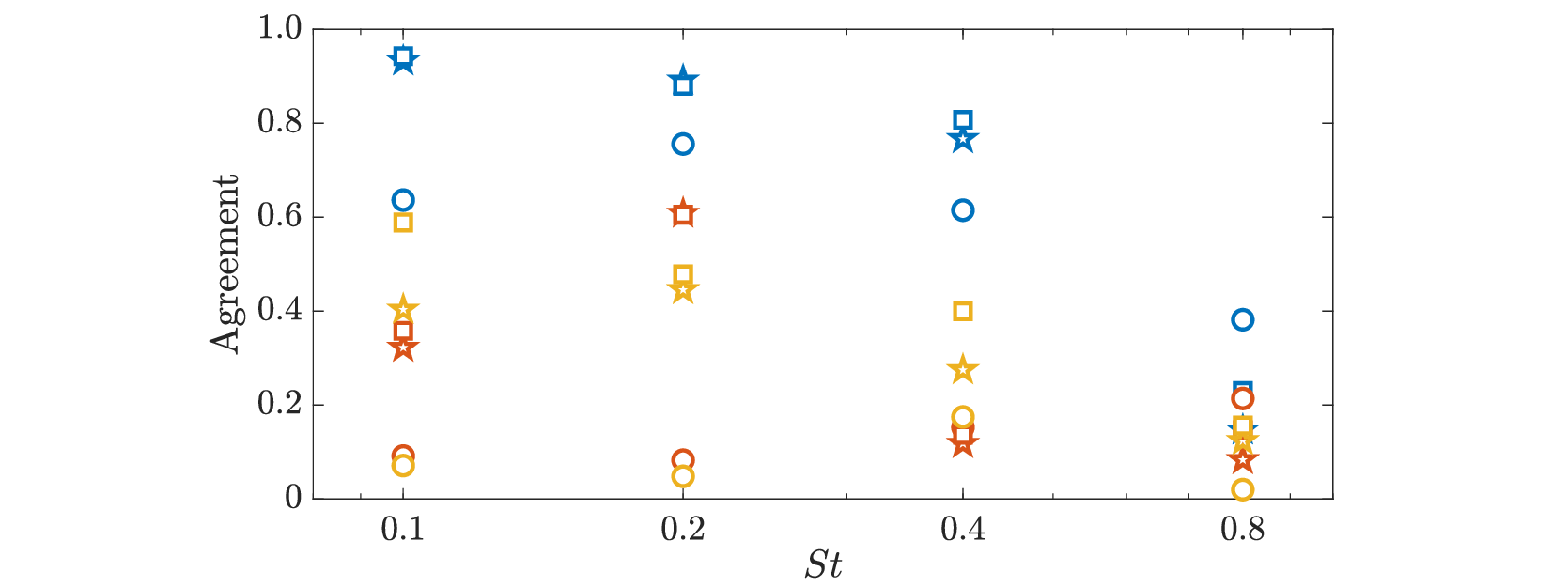}}
	\caption{Quantitative agreement (1 means complete agreement and 0 means no agreement) between the resolvent modes (calculated based on the $\nu$ (circles), $\nu_w$ (stars) and $\nu_b$ (squares) models) and the SPOD modes. The results for the first three modes at $m=1$ and at $St=0.1$, $0.2$, $0.4$ and $0.8$ are presented. The blue, red and yellow symbols represent modes 1, 2 and 3 respectively.}
	\label{fig_wakeRes_alignment_RA-SPOD}
\end{figure}

In Fig.~\ref{fig_wakeRes_mode_SPOD-RA_viscEffect}, we compare the effect of different eddy-viscosity models on the mode shapes. The resolvent modes for both the $\nu_w$- and $\nu_b$-models agree well with the SPOD modes. However, the resolvent modes are relatively concentrated near the centerline for the results from $\nu$-model. Such magnitude peaks are not observed in the SPOD modes. To quantitatively assess the mode agreement, we define an agreement measure as the dot product between the conjugate of the resolvent and SPOD modes~\cite{pickering2021optimal} in the far-wake region ($5\leq x/D \leq 10$ and $0\leq r/D \leq 3$). In this measure, values of approximately 0.4 or greater show qualitative agreement, whereas values less than 0.4 have little visual similarity~\cite{pickering2021optimal}. In Fig.~\ref{fig_wakeRes_alignment_RA-SPOD}, we examine the mode agreement when different eddy-viscosity models are used. At $St=0.1$, $0.2$, and $St=0.4$, the $\nu_w$- and $\nu_b$-models exhibit significantly better agreement than the $\nu$-model. The inconsistent agreement at $St=0.8$ may be attributed to the inability of the resolvent analysis to capture turbulence generation that is not dominated by convective instabilities. These findings indicate that for accurate prediction of the far-wake energetic structures, the resolvent operator needs to be augmented with an appropriate eddy-viscosity model.

For the frequency at which the optimal resolvent gain occurs, i.e., $St=0.2$, Fig.~\ref{fig_wakeRes_alignment_RA-SPOD} shows that the $\nu_w$-model only slightly improves the mode agreement compared to the $\nu_b$-model. The $\nu_w$-model also exhibits slightly better performance than the $\nu_b$-model in both the resolvent gain spectra (see Fig.~\ref{fig_wakeRes_spectra_viscEffect}) and the mode agreement (see Fig.~\ref{fig_wakeRes_alignment_RA-SPOD}) due to its ability to account for spatial variability in the eddy-viscosity field. For all dominant frequencies of the resolvent gains, such as $St=0.1$, $0.2$, and $0.4$, both the $\nu_w$- and $\nu_b$-models show good agreement (up to approximately 95$\%$ for leading modes and up to approximately 60$\%$ for higher-rank modes). We therefore consider resolvent analysis results not to be overly sensitive to the eddy viscosity profiles.

\section{Conclusions}\label{sec-wakeRes-conc}

We have used resolvent analysis to predict the energetic structures in the far-wake of a wind turbine operating in a neutral atmospheric boundary layer. The flow was numerically simulated with LES, which generated the mean flow and eddy viscosity profiles required by the resolvent analysis. We restricted the analysis to the spanwise plane passing through the turbine center with the approximation of axisymmetric wake flow, which is consistent with most engineering wake models. To validate our results, we compared the energetic modes from resolvent analysis with those obtained through SPOD. We found that it is necessary to include the eddy viscosity in the resolvent operator to be able to accurately account for small-scale turbulence, while the axisymmetric wake flow approximation is satisfactory.

The resolvent analysis shows that the radial forcing component exerts the most significant influence on TKE generation in the wake flow relative to its magnitude in the inflow fluctuations. The resolvent gain is found to peak at $St \approx 0.2$ and $|m| = 1$ and the response mode is the most energetic in the high mean shear region that coincides with the maximum TKE region. These findings reveal a dominant role of convective instabilities in generating far-wake turbulence, which is consistent with the existing literature. Additionally, resolvent analysis also uncovers a potential role of the non-modal Orr mechanism in enhancing TKE generation, particularly at low frequencies. This is evidenced from the shape of the optimal forcing modes, which are tilted against the mean shear.

%
We find that resolvent analysis can predict the spectral characteristics and mode shapes in the far-wake region well enough for reasonable agreement with those obtained through SPOD. Specifically, Fig.~\ref{fig_wakeRes_spectra_RA-SPOD} shows that multiplying the optimal gain with the inflow TKE spectrum reproduces well the wake TKE spectra. The resolvent modes can be reconstructed to establish a reduced-order model~\cite{gomez2016reduced,pickering2021resolvent} that predicts time-varying characteristics of the far-wake flow. Formulating the resolvent operator only requires the mean flow and eddy viscosity fields as inputs, which can be obtained from existing engineering wake models~\cite{larsen2008wake, jonkman2017development}, and captures the energetic modes despite the axisymmetric wake flow approximation. The resolvent-based model, therefore, can potentially be integrated with existing wake models for accurate prediction of TKE generation in the far wakes.

The present study also sets the stage for further improvements in the application of resolvent analysis to wind turbine wakes, which can be in two aspects. First, the modelling of the nonlinear term can be improved, such as in~\cite{pickering2021optimal, gupta2021linear, wu2023composition,madhusudanan2022navier}. This could be particularly important for flows with thermal stratification, which may need the consideration of energy equation~\cite{madhusudanan2022navier}, and changes in the length scale of atmospheric turbulence. For instance, our conclusion on the role of eddy viscosity may not be valid in stably stratified ABL, which has lower levels of turbulence, and in unstably stratified turbulence, which has strong convection-induced rolls. The second aspect is related to the non-axisymmetric effects. Our results show that the azimuthal modes from SPOD and resolvent analysis do not match well. This indicates the need for relaxing the approximation of axisymmetry to account for three-dimensional effects, such as those arise via the interaction between the turbine wake and the boundary layer flows.


\begin{acknowledgments}
This work was supported by the National Natural Science Foundation of China (Grant Nos. 12225204, 12002147 and 12050410247), the Department of Science and Technology of Guangdong Province (Grant Nos. 2019B21203001 and 2020B1212030001) and the Research Grants Council of Hong Kong (Grant No. 16200220). We would like to thank Simon Watson from Delft University of Technology for feedback that helped us improve the manuscript. We acknowledge support from the Centers for Mechanical Engineering Research and Education at MIT and SUSTech, as well as from the Center for Computational Science and Engineering at SUSTech.
\end{acknowledgments}

\appendix
\section{Two-dimensional resolvent analysis of turbine wake flow in the vertical plane}\label{apdx_SPODXZ}

\begin{figure}[htbp]
	\centerline{\includegraphics[width=1\textwidth]{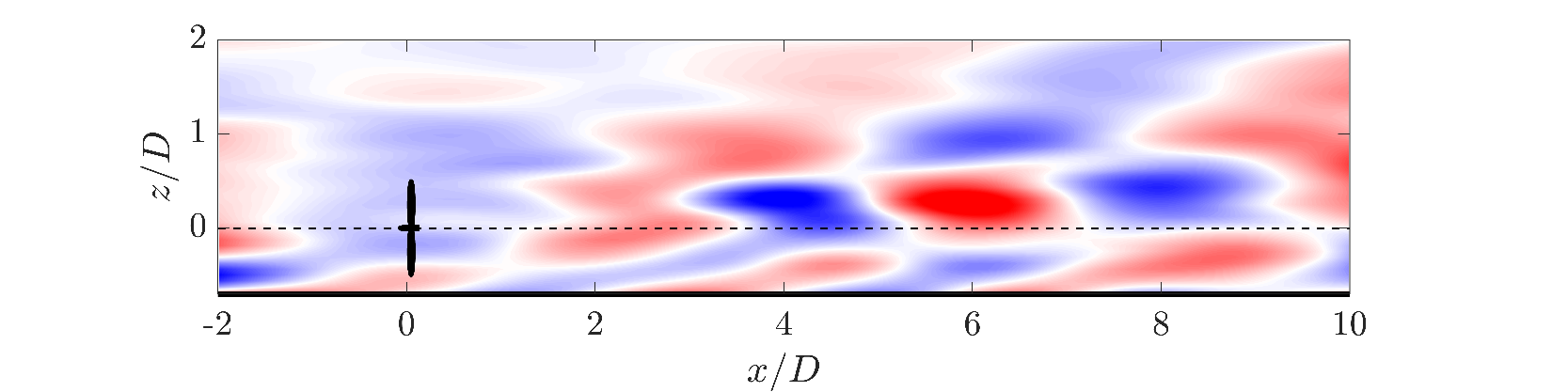}}
	\caption{The $x$-velocity component of the dominant SPOD mode at $St=0.2$. The mode is computed in the $x$-$z$ plane passing the turbine center and normalized by the maximum value over all velocity components and frequencies. The dashed line indicates the turbine centerline and the bottom solid line indicates the ground.}
	\label{fig_wakeRes_mode_SPOD_XZ}
\end{figure}

The wake flow obtained from the LES data, influenced by atmospheric shear and turbine rotation, exhibits full three-dimensional characteristics without any homogeneous directions. However, existing dynamic wake models~\cite{larsen2008wake,jonkman2017development} commonly assume axisymmetric wake flow, where the azimuthal direction is considered homogeneous. Additionally, it has been observed that with the axisymmetry approximation, resolvent analysis conducted on the $x$-$y$ plane (as shown in the resolvent mode at $St=0.2$ in Fig.~\ref{fig_wakeRes_mode_SPOD-RA_freq}) along the turbine centerline can effectively predict the far-wake energetic structures identified through SPOD in both the $x$-$y$ plane (as shown in the SPOD mode at $St=0.2$ in Fig.\ref{fig_wakeRes_mode_SPOD-RA_freq}) and the $x$-$z$ plane (as shown in the mode above the turbine centerline in Fig.~\ref{fig_wakeRes_mode_SPOD_XZ}).

Resolvent analysis can also be performed on the $x$-$z$ plane along the turbine centerline, assuming spanwise homogeneity. Such analysis is expected to capture the energetic structures concentrated in both the wake region and the high-shear region near the ground (as shown in the mode beneath the turbine centerline in Fig.~\ref{fig_wakeRes_mode_SPOD_XZ}). These structures arise due to the presence of both the turbine and the ground, leading to dynamics that differ from those of the far-wake region. In the present study, we focus on the dynamics of the far-wake region alone. The interaction between the far-wake and the wall turbulence, however, is beyond the scope of the present study.

\begin{thebibliography}{85}%
\makeatletter
\providecommand \@ifxundefined [1]{%
 \@ifx{#1\undefined}
}%
\providecommand \@ifnum [1]{%
 \ifnum #1\expandafter \@firstoftwo
 \else \expandafter \@secondoftwo
 \fi
}%
\providecommand \@ifx [1]{%
 \ifx #1\expandafter \@firstoftwo
 \else \expandafter \@secondoftwo
 \fi
}%
\providecommand \natexlab [1]{#1}%
\providecommand \enquote  [1]{``#1''}%
\providecommand \bibnamefont  [1]{#1}%
\providecommand \bibfnamefont [1]{#1}%
\providecommand \citenamefont [1]{#1}%
\providecommand \href@noop [0]{\@secondoftwo}%
\providecommand \href [0]{\begingroup \@sanitize@url \@href}%
\providecommand \@href[1]{\@@startlink{#1}\@@href}%
\providecommand \@@href[1]{\endgroup#1\@@endlink}%
\providecommand \@sanitize@url [0]{\catcode `\\12\catcode `\$12\catcode
  `\&12\catcode `\#12\catcode `\^12\catcode `\_12\catcode `\%12\relax}%
\providecommand \@@startlink[1]{}%
\providecommand \@@endlink[0]{}%
\providecommand \url  [0]{\begingroup\@sanitize@url \@url }%
\providecommand \@url [1]{\endgroup\@href {#1}{\urlprefix }}%
\providecommand \urlprefix  [0]{URL }%
\providecommand \Eprint [0]{\href }%
\providecommand \doibase [0]{https://doi.org/}%
\providecommand \selectlanguage [0]{\@gobble}%
\providecommand \bibinfo  [0]{\@secondoftwo}%
\providecommand \bibfield  [0]{\@secondoftwo}%
\providecommand \translation [1]{[#1]}%
\providecommand \BibitemOpen [0]{}%
\providecommand \bibitemStop [0]{}%
\providecommand \bibitemNoStop [0]{.\EOS\space}%
\providecommand \EOS [0]{\spacefactor3000\relax}%
\providecommand \BibitemShut  [1]{\csname bibitem#1\endcsname}%
\let\auto@bib@innerbib\@empty
\bibitem [{\citenamefont {Port{\'e}-Agel}\ \emph {et~al.}(2020)\citenamefont
  {Port{\'e}-Agel}, \citenamefont {Bastankhah},\ and\ \citenamefont
  {Shamsoddin}}]{porte2020wind}%
  \BibitemOpen
  \bibfield  {author} {\bibinfo {author} {\bibfnamefont {F.}~\bibnamefont
  {Port{\'e}-Agel}}, \bibinfo {author} {\bibfnamefont {M.}~\bibnamefont
  {Bastankhah}},\ and\ \bibinfo {author} {\bibfnamefont {S.}~\bibnamefont
  {Shamsoddin}},\ }\bibfield  {title} {\bibinfo {title} {Wind-turbine and
  wind-farm flows: A review},\ }\href@noop {} {\bibfield  {journal} {\bibinfo
  {journal} {Boundary-layer meteorology}\ }\textbf {\bibinfo {volume} {174}},\
  \bibinfo {pages} {1} (\bibinfo {year} {2020})}\BibitemShut {NoStop}%
\bibitem [{\citenamefont {Stevens}\ and\ \citenamefont
  {Meneveau}(2017)}]{stevens2017flow}%
  \BibitemOpen
  \bibfield  {author} {\bibinfo {author} {\bibfnamefont {R.~J.}\ \bibnamefont
  {Stevens}}\ and\ \bibinfo {author} {\bibfnamefont {C.}~\bibnamefont
  {Meneveau}},\ }\bibfield  {title} {\bibinfo {title} {Flow structure and
  turbulence in wind farms},\ }\href@noop {} {\bibfield  {journal} {\bibinfo
  {journal} {Annual review of fluid mechanics}\ }\textbf {\bibinfo {volume}
  {49}},\ \bibinfo {pages} {311} (\bibinfo {year} {2017})}\BibitemShut
  {NoStop}%
\bibitem [{\citenamefont {Yang}\ and\ \citenamefont
  {Sotiropoulos}(2019)}]{yang2019review}%
  \BibitemOpen
  \bibfield  {author} {\bibinfo {author} {\bibfnamefont {X.}~\bibnamefont
  {Yang}}\ and\ \bibinfo {author} {\bibfnamefont {F.}~\bibnamefont
  {Sotiropoulos}},\ }\bibfield  {title} {\bibinfo {title} {A review on the
  meandering of wind turbine wakes},\ }\href@noop {} {\bibfield  {journal}
  {\bibinfo  {journal} {Energies}\ }\textbf {\bibinfo {volume} {12}},\ \bibinfo
  {pages} {4725} (\bibinfo {year} {2019})}\BibitemShut {NoStop}%
\bibitem [{\citenamefont {Veers}\ \emph {et~al.}(2019)\citenamefont {Veers},
  \citenamefont {Dykes}, \citenamefont {Lantz}, \citenamefont {Barth},
  \citenamefont {Bottasso}, \citenamefont {Carlson}, \citenamefont {Clifton},
  \citenamefont {Green}, \citenamefont {Green}, \citenamefont {Holttinen} \emph
  {et~al.}}]{veers2019grand}%
  \BibitemOpen
  \bibfield  {author} {\bibinfo {author} {\bibfnamefont {P.}~\bibnamefont
  {Veers}}, \bibinfo {author} {\bibfnamefont {K.}~\bibnamefont {Dykes}},
  \bibinfo {author} {\bibfnamefont {E.}~\bibnamefont {Lantz}}, \bibinfo
  {author} {\bibfnamefont {S.}~\bibnamefont {Barth}}, \bibinfo {author}
  {\bibfnamefont {C.~L.}\ \bibnamefont {Bottasso}}, \bibinfo {author}
  {\bibfnamefont {O.}~\bibnamefont {Carlson}}, \bibinfo {author} {\bibfnamefont
  {A.}~\bibnamefont {Clifton}}, \bibinfo {author} {\bibfnamefont
  {J.}~\bibnamefont {Green}}, \bibinfo {author} {\bibfnamefont
  {P.}~\bibnamefont {Green}}, \bibinfo {author} {\bibfnamefont
  {H.}~\bibnamefont {Holttinen}}, \emph {et~al.},\ }\bibfield  {title}
  {\bibinfo {title} {Grand challenges in the science of wind energy},\
  }\href@noop {} {\bibfield  {journal} {\bibinfo  {journal} {Science}\ }\textbf
  {\bibinfo {volume} {366}},\ \bibinfo {pages} {eaau2027} (\bibinfo {year}
  {2019})}\BibitemShut {NoStop}%
\bibitem [{\citenamefont {Larsen}\ \emph {et~al.}(2008)\citenamefont {Larsen},
  \citenamefont {Madsen}, \citenamefont {Thomsen},\ and\ \citenamefont
  {Larsen}}]{larsen2008wake}%
  \BibitemOpen
  \bibfield  {author} {\bibinfo {author} {\bibfnamefont {G.~C.}\ \bibnamefont
  {Larsen}}, \bibinfo {author} {\bibfnamefont {H.~A.}\ \bibnamefont {Madsen}},
  \bibinfo {author} {\bibfnamefont {K.}~\bibnamefont {Thomsen}},\ and\ \bibinfo
  {author} {\bibfnamefont {T.~J.}\ \bibnamefont {Larsen}},\ }\bibfield  {title}
  {\bibinfo {title} {Wake meandering: a pragmatic approach},\ }\href@noop {}
  {\bibfield  {journal} {\bibinfo  {journal} {Wind Energy: An International
  Journal for Progress and Applications in Wind Power Conversion Technology}\
  }\textbf {\bibinfo {volume} {11}},\ \bibinfo {pages} {377} (\bibinfo {year}
  {2008})}\BibitemShut {NoStop}%
\bibitem [{\citenamefont {Bastine}\ \emph {et~al.}(2015)\citenamefont
  {Bastine}, \citenamefont {Witha}, \citenamefont {W{\"a}chter},\ and\
  \citenamefont {Peinke}}]{bastine2015towards}%
  \BibitemOpen
  \bibfield  {author} {\bibinfo {author} {\bibfnamefont {D.}~\bibnamefont
  {Bastine}}, \bibinfo {author} {\bibfnamefont {B.}~\bibnamefont {Witha}},
  \bibinfo {author} {\bibfnamefont {M.}~\bibnamefont {W{\"a}chter}},\ and\
  \bibinfo {author} {\bibfnamefont {J.}~\bibnamefont {Peinke}},\ }\bibfield
  {title} {\bibinfo {title} {Towards a simplified dynamic wake model using pod
  analysis},\ }\href@noop {} {\bibfield  {journal} {\bibinfo  {journal}
  {Energies}\ }\textbf {\bibinfo {volume} {8}},\ \bibinfo {pages} {895}
  (\bibinfo {year} {2015})}\BibitemShut {NoStop}%
\bibitem [{\citenamefont {Th{\o}gersen}\ \emph {et~al.}(2017)\citenamefont
  {Th{\o}gersen}, \citenamefont {Tranberg}, \citenamefont {Herp},\ and\
  \citenamefont {Greiner}}]{thogersen2017statistical}%
  \BibitemOpen
  \bibfield  {author} {\bibinfo {author} {\bibfnamefont {E.}~\bibnamefont
  {Th{\o}gersen}}, \bibinfo {author} {\bibfnamefont {B.}~\bibnamefont
  {Tranberg}}, \bibinfo {author} {\bibfnamefont {J.}~\bibnamefont {Herp}},\
  and\ \bibinfo {author} {\bibfnamefont {M.}~\bibnamefont {Greiner}},\
  }\bibfield  {title} {\bibinfo {title} {Statistical meandering wake model and
  its application to yaw-angle optimisation of wind farms},\ }in\ \href@noop {}
  {\emph {\bibinfo {booktitle} {Journal of Physics: Conference Series}}},\
  Vol.\ \bibinfo {volume} {854}\ (\bibinfo {organization} {IOP Publishing},\
  \bibinfo {year} {2017})\ p.\ \bibinfo {pages} {012017}\BibitemShut {NoStop}%
\bibitem [{\citenamefont {Mao}\ and\ \citenamefont
  {S{\o}rensen}(2018)}]{mao2018far}%
  \BibitemOpen
  \bibfield  {author} {\bibinfo {author} {\bibfnamefont {X.}~\bibnamefont
  {Mao}}\ and\ \bibinfo {author} {\bibfnamefont {J.}~\bibnamefont
  {S{\o}rensen}},\ }\bibfield  {title} {\bibinfo {title} {Far-wake meandering
  induced by atmospheric eddies in flow past a wind turbine},\ }\href@noop {}
  {\bibfield  {journal} {\bibinfo  {journal} {Journal of Fluid Mechanics}\
  }\textbf {\bibinfo {volume} {846}},\ \bibinfo {pages} {190} (\bibinfo {year}
  {2018})}\BibitemShut {NoStop}%
\bibitem [{\citenamefont {Gupta}\ and\ \citenamefont
  {Wan}(2019)}]{gupta2019low}%
  \BibitemOpen
  \bibfield  {author} {\bibinfo {author} {\bibfnamefont {V.}~\bibnamefont
  {Gupta}}\ and\ \bibinfo {author} {\bibfnamefont {M.}~\bibnamefont {Wan}},\
  }\bibfield  {title} {\bibinfo {title} {Low-order modelling of wake meandering
  behind turbines},\ }\href@noop {} {\bibfield  {journal} {\bibinfo  {journal}
  {Journal of Fluid Mechanics}\ }\textbf {\bibinfo {volume} {877}},\ \bibinfo
  {pages} {534} (\bibinfo {year} {2019})}\BibitemShut {NoStop}%
\bibitem [{\citenamefont {Shapiro}\ \emph {et~al.}(2017)\citenamefont
  {Shapiro}, \citenamefont {Bauweraerts}, \citenamefont {Meyers}, \citenamefont
  {Meneveau},\ and\ \citenamefont {Gayme}}]{shapiro2017model}%
  \BibitemOpen
  \bibfield  {author} {\bibinfo {author} {\bibfnamefont {C.~R.}\ \bibnamefont
  {Shapiro}}, \bibinfo {author} {\bibfnamefont {P.}~\bibnamefont
  {Bauweraerts}}, \bibinfo {author} {\bibfnamefont {J.}~\bibnamefont {Meyers}},
  \bibinfo {author} {\bibfnamefont {C.}~\bibnamefont {Meneveau}},\ and\
  \bibinfo {author} {\bibfnamefont {D.~F.}\ \bibnamefont {Gayme}},\ }\bibfield
  {title} {\bibinfo {title} {Model-based receding horizon control of wind farms
  for secondary frequency regulation},\ }\href@noop {} {\bibfield  {journal}
  {\bibinfo  {journal} {Wind Energy}\ }\textbf {\bibinfo {volume} {20}},\
  \bibinfo {pages} {1261} (\bibinfo {year} {2017})}\BibitemShut {NoStop}%
\bibitem [{\citenamefont {Gebraad}\ \emph {et~al.}(2016)\citenamefont
  {Gebraad}, \citenamefont {Teeuwisse}, \citenamefont {Van~Wingerden},
  \citenamefont {Fleming}, \citenamefont {Ruben}, \citenamefont {Marden},\ and\
  \citenamefont {Pao}}]{gebraad2016wind}%
  \BibitemOpen
  \bibfield  {author} {\bibinfo {author} {\bibfnamefont {P.~M.}\ \bibnamefont
  {Gebraad}}, \bibinfo {author} {\bibfnamefont {F.~W.}\ \bibnamefont
  {Teeuwisse}}, \bibinfo {author} {\bibfnamefont {J.}~\bibnamefont
  {Van~Wingerden}}, \bibinfo {author} {\bibfnamefont {P.~A.}\ \bibnamefont
  {Fleming}}, \bibinfo {author} {\bibfnamefont {S.~D.}\ \bibnamefont {Ruben}},
  \bibinfo {author} {\bibfnamefont {J.~R.}\ \bibnamefont {Marden}},\ and\
  \bibinfo {author} {\bibfnamefont {L.~Y.}\ \bibnamefont {Pao}},\ }\bibfield
  {title} {\bibinfo {title} {Wind plant power optimization through yaw control
  using a parametric model for wake effects—a cfd simulation study},\
  }\href@noop {} {\bibfield  {journal} {\bibinfo  {journal} {Wind Energy}\
  }\textbf {\bibinfo {volume} {19}},\ \bibinfo {pages} {95} (\bibinfo {year}
  {2016})}\BibitemShut {NoStop}%
\bibitem [{\citenamefont {Vermeer}\ \emph {et~al.}(2003)\citenamefont
  {Vermeer}, \citenamefont {S{\o}rensen},\ and\ \citenamefont
  {Crespo}}]{vermeer2003wind}%
  \BibitemOpen
  \bibfield  {author} {\bibinfo {author} {\bibfnamefont {L.}~\bibnamefont
  {Vermeer}}, \bibinfo {author} {\bibfnamefont {J.~N.}\ \bibnamefont
  {S{\o}rensen}},\ and\ \bibinfo {author} {\bibfnamefont {A.}~\bibnamefont
  {Crespo}},\ }\bibfield  {title} {\bibinfo {title} {Wind turbine wake
  aerodynamics},\ }\href@noop {} {\bibfield  {journal} {\bibinfo  {journal}
  {Progress in aerospace sciences}\ }\textbf {\bibinfo {volume} {39}},\
  \bibinfo {pages} {467} (\bibinfo {year} {2003})}\BibitemShut {NoStop}%
\bibitem [{\citenamefont {Han}\ \emph {et~al.}(2018)\citenamefont {Han},
  \citenamefont {Liu}, \citenamefont {Xu},\ and\ \citenamefont
  {Shen}}]{han2018atmospheric}%
  \BibitemOpen
  \bibfield  {author} {\bibinfo {author} {\bibfnamefont {X.}~\bibnamefont
  {Han}}, \bibinfo {author} {\bibfnamefont {D.}~\bibnamefont {Liu}}, \bibinfo
  {author} {\bibfnamefont {C.}~\bibnamefont {Xu}},\ and\ \bibinfo {author}
  {\bibfnamefont {W.~Z.}\ \bibnamefont {Shen}},\ }\bibfield  {title} {\bibinfo
  {title} {Atmospheric stability and topography effects on wind turbine
  performance and wake properties in complex terrain},\ }\href@noop {}
  {\bibfield  {journal} {\bibinfo  {journal} {Renewable energy}\ }\textbf
  {\bibinfo {volume} {126}},\ \bibinfo {pages} {640} (\bibinfo {year}
  {2018})}\BibitemShut {NoStop}%
\bibitem [{\citenamefont {Heisel}\ \emph {et~al.}(2018)\citenamefont {Heisel},
  \citenamefont {Hong},\ and\ \citenamefont {Guala}}]{heisel2018spectral}%
  \BibitemOpen
  \bibfield  {author} {\bibinfo {author} {\bibfnamefont {M.}~\bibnamefont
  {Heisel}}, \bibinfo {author} {\bibfnamefont {J.}~\bibnamefont {Hong}},\ and\
  \bibinfo {author} {\bibfnamefont {M.}~\bibnamefont {Guala}},\ }\bibfield
  {title} {\bibinfo {title} {The spectral signature of wind turbine wake
  meandering: A wind tunnel and field-scale study},\ }\href@noop {} {\bibfield
  {journal} {\bibinfo  {journal} {Wind Energy}\ }\textbf {\bibinfo {volume}
  {21}},\ \bibinfo {pages} {715} (\bibinfo {year} {2018})}\BibitemShut
  {NoStop}%
\bibitem [{\citenamefont {Coudou}\ \emph {et~al.}(2018)\citenamefont {Coudou},
  \citenamefont {Buckingham}, \citenamefont {Bricteux},\ and\ \citenamefont
  {van Beeck}}]{coudou2018experimental}%
  \BibitemOpen
  \bibfield  {author} {\bibinfo {author} {\bibfnamefont {N.}~\bibnamefont
  {Coudou}}, \bibinfo {author} {\bibfnamefont {S.}~\bibnamefont {Buckingham}},
  \bibinfo {author} {\bibfnamefont {L.}~\bibnamefont {Bricteux}},\ and\
  \bibinfo {author} {\bibfnamefont {J.}~\bibnamefont {van Beeck}},\ }\bibfield
  {title} {\bibinfo {title} {Experimental study on the wake meandering within a
  scale model wind farm subject to a wind-tunnel flow simulating an atmospheric
  boundary layer},\ }\href@noop {} {\bibfield  {journal} {\bibinfo  {journal}
  {Boundary-layer meteorology}\ }\textbf {\bibinfo {volume} {167}},\ \bibinfo
  {pages} {77} (\bibinfo {year} {2018})}\BibitemShut {NoStop}%
\bibitem [{\citenamefont {Foti}\ \emph {et~al.}(2018)\citenamefont {Foti},
  \citenamefont {Yang},\ and\ \citenamefont
  {Sotiropoulos}}]{foti2018similarity}%
  \BibitemOpen
  \bibfield  {author} {\bibinfo {author} {\bibfnamefont {D.}~\bibnamefont
  {Foti}}, \bibinfo {author} {\bibfnamefont {X.}~\bibnamefont {Yang}},\ and\
  \bibinfo {author} {\bibfnamefont {F.}~\bibnamefont {Sotiropoulos}},\
  }\bibfield  {title} {\bibinfo {title} {Similarity of wake meandering for
  different wind turbine designs for different scales},\ }\href@noop {}
  {\bibfield  {journal} {\bibinfo  {journal} {Journal of Fluid Mechanics}\
  }\textbf {\bibinfo {volume} {842}},\ \bibinfo {pages} {5} (\bibinfo {year}
  {2018})}\BibitemShut {NoStop}%
\bibitem [{\citenamefont {Feng}\ \emph {et~al.}(2022)\citenamefont {Feng},
  \citenamefont {Li}, \citenamefont {Gupta},\ and\ \citenamefont
  {Wan}}]{feng2022componentwise}%
  \BibitemOpen
  \bibfield  {author} {\bibinfo {author} {\bibfnamefont {D.}~\bibnamefont
  {Feng}}, \bibinfo {author} {\bibfnamefont {L.~K.}\ \bibnamefont {Li}},
  \bibinfo {author} {\bibfnamefont {V.}~\bibnamefont {Gupta}},\ and\ \bibinfo
  {author} {\bibfnamefont {M.}~\bibnamefont {Wan}},\ }\bibfield  {title}
  {\bibinfo {title} {Componentwise influence of upstream turbulence on the
  far-wake dynamics of wind turbines},\ }\href@noop {} {\bibfield  {journal}
  {\bibinfo  {journal} {Renewable Energy}\ }\textbf {\bibinfo {volume} {200}},\
  \bibinfo {pages} {1081} (\bibinfo {year} {2022})}\BibitemShut {NoStop}%
\bibitem [{\citenamefont {Gebraad}\ and\ \citenamefont
  {Van~Wingerden}(2014)}]{gebraad2014control}%
  \BibitemOpen
  \bibfield  {author} {\bibinfo {author} {\bibfnamefont {P.~M.}\ \bibnamefont
  {Gebraad}}\ and\ \bibinfo {author} {\bibfnamefont {J.}~\bibnamefont
  {Van~Wingerden}},\ }\bibfield  {title} {\bibinfo {title} {A control-oriented
  dynamic model for wakes in wind plants},\ }in\ \href@noop {} {\emph {\bibinfo
  {booktitle} {Journal of Physics: Conference Series}}},\ Vol.\ \bibinfo
  {volume} {524}\ (\bibinfo {organization} {IOP Publishing},\ \bibinfo {year}
  {2014})\ p.\ \bibinfo {pages} {012186}\BibitemShut {NoStop}%
\bibitem [{\citenamefont {Braunbehrens}\ and\ \citenamefont
  {Segalini}(2019)}]{braunbehrens2019statistical}%
  \BibitemOpen
  \bibfield  {author} {\bibinfo {author} {\bibfnamefont {R.}~\bibnamefont
  {Braunbehrens}}\ and\ \bibinfo {author} {\bibfnamefont {A.}~\bibnamefont
  {Segalini}},\ }\bibfield  {title} {\bibinfo {title} {A statistical model for
  wake meandering behind wind turbines},\ }\href@noop {} {\bibfield  {journal}
  {\bibinfo  {journal} {Journal of Wind Engineering and Industrial
  Aerodynamics}\ }\textbf {\bibinfo {volume} {193}},\ \bibinfo {pages} {103954}
  (\bibinfo {year} {2019})}\BibitemShut {NoStop}%
\bibitem [{\citenamefont {Debnath}\ \emph {et~al.}(2017)\citenamefont
  {Debnath}, \citenamefont {Santoni}, \citenamefont {Leonardi},\ and\
  \citenamefont {Iungo}}]{debnath2017towards}%
  \BibitemOpen
  \bibfield  {author} {\bibinfo {author} {\bibfnamefont {M.}~\bibnamefont
  {Debnath}}, \bibinfo {author} {\bibfnamefont {C.}~\bibnamefont {Santoni}},
  \bibinfo {author} {\bibfnamefont {S.}~\bibnamefont {Leonardi}},\ and\
  \bibinfo {author} {\bibfnamefont {G.~V.}\ \bibnamefont {Iungo}},\ }\bibfield
  {title} {\bibinfo {title} {Towards reduced order modelling for predicting the
  dynamics of coherent vorticity structures within wind turbine wakes},\
  }\href@noop {} {\bibfield  {journal} {\bibinfo  {journal} {Philosophical
  Transactions of the Royal Society A: Mathematical, Physical and Engineering
  Sciences}\ }\textbf {\bibinfo {volume} {375}},\ \bibinfo {pages} {20160108}
  (\bibinfo {year} {2017})}\BibitemShut {NoStop}%
\bibitem [{\citenamefont {Bastine}\ \emph {et~al.}(2018)\citenamefont
  {Bastine}, \citenamefont {Vollmer}, \citenamefont {W{\"a}chter},\ and\
  \citenamefont {Peinke}}]{bastine2018stochastic}%
  \BibitemOpen
  \bibfield  {author} {\bibinfo {author} {\bibfnamefont {D.}~\bibnamefont
  {Bastine}}, \bibinfo {author} {\bibfnamefont {L.}~\bibnamefont {Vollmer}},
  \bibinfo {author} {\bibfnamefont {M.}~\bibnamefont {W{\"a}chter}},\ and\
  \bibinfo {author} {\bibfnamefont {J.}~\bibnamefont {Peinke}},\ }\bibfield
  {title} {\bibinfo {title} {Stochastic wake modelling based on pod analysis},\
  }\href@noop {} {\bibfield  {journal} {\bibinfo  {journal} {Energies}\
  }\textbf {\bibinfo {volume} {11}},\ \bibinfo {pages} {612} (\bibinfo {year}
  {2018})}\BibitemShut {NoStop}%
\bibitem [{\citenamefont {Iungo}\ \emph {et~al.}(2015)\citenamefont {Iungo},
  \citenamefont {Santoni-Ortiz}, \citenamefont {Abkar}, \citenamefont
  {Port{\'e}-Agel}, \citenamefont {Rotea},\ and\ \citenamefont
  {Leonardi}}]{iungo2015data}%
  \BibitemOpen
  \bibfield  {author} {\bibinfo {author} {\bibfnamefont {G.~V.}\ \bibnamefont
  {Iungo}}, \bibinfo {author} {\bibfnamefont {C.}~\bibnamefont
  {Santoni-Ortiz}}, \bibinfo {author} {\bibfnamefont {M.}~\bibnamefont
  {Abkar}}, \bibinfo {author} {\bibfnamefont {F.}~\bibnamefont
  {Port{\'e}-Agel}}, \bibinfo {author} {\bibfnamefont {M.~A.}\ \bibnamefont
  {Rotea}},\ and\ \bibinfo {author} {\bibfnamefont {S.}~\bibnamefont
  {Leonardi}},\ }\bibfield  {title} {\bibinfo {title} {Data-driven reduced
  order model for prediction of wind turbine wakes},\ }in\ \href@noop {} {\emph
  {\bibinfo {booktitle} {Journal of Physics: Conference Series}}},\ Vol.\
  \bibinfo {volume} {625}\ (\bibinfo {organization} {IOP Publishing},\ \bibinfo
  {year} {2015})\ p.\ \bibinfo {pages} {012009}\BibitemShut {NoStop}%
\bibitem [{\citenamefont {Iungo}\ \emph {et~al.}(2022)\citenamefont {Iungo},
  \citenamefont {Maulik}, \citenamefont {Renganathan},\ and\ \citenamefont
  {Letizia}}]{iungo2022machine}%
  \BibitemOpen
  \bibfield  {author} {\bibinfo {author} {\bibfnamefont {G.~V.}\ \bibnamefont
  {Iungo}}, \bibinfo {author} {\bibfnamefont {R.}~\bibnamefont {Maulik}},
  \bibinfo {author} {\bibfnamefont {S.~A.}\ \bibnamefont {Renganathan}},\ and\
  \bibinfo {author} {\bibfnamefont {S.}~\bibnamefont {Letizia}},\ }\bibfield
  {title} {\bibinfo {title} {Machine-learning identification of the variability
  of mean velocity and turbulence intensity for wakes generated by onshore wind
  turbines: Cluster analysis of wind lidar measurements},\ }\href@noop {}
  {\bibfield  {journal} {\bibinfo  {journal} {Journal of Renewable and
  Sustainable Energy}\ }\textbf {\bibinfo {volume} {14}} (\bibinfo {year}
  {2022})}\BibitemShut {NoStop}%
\bibitem [{\citenamefont {Ali}\ \emph {et~al.}(2021)\citenamefont {Ali},
  \citenamefont {Calaf},\ and\ \citenamefont {Cal}}]{ali2021clustering}%
  \BibitemOpen
  \bibfield  {author} {\bibinfo {author} {\bibfnamefont {N.}~\bibnamefont
  {Ali}}, \bibinfo {author} {\bibfnamefont {M.}~\bibnamefont {Calaf}},\ and\
  \bibinfo {author} {\bibfnamefont {R.~B.}\ \bibnamefont {Cal}},\ }\bibfield
  {title} {\bibinfo {title} {Clustering sparse sensor placement identification
  and deep learning based forecasting for wind turbine wakes},\ }\href@noop {}
  {\bibfield  {journal} {\bibinfo  {journal} {Journal of Renewable and
  Sustainable Energy}\ }\textbf {\bibinfo {volume} {13}} (\bibinfo {year}
  {2021})}\BibitemShut {NoStop}%
\bibitem [{\citenamefont {Moon}\ \emph {et~al.}(2017)\citenamefont {Moon},
  \citenamefont {Manuel}, \citenamefont {Churchfield}, \citenamefont {Lee},\
  and\ \citenamefont {Veers}}]{moon2017toward}%
  \BibitemOpen
  \bibfield  {author} {\bibinfo {author} {\bibfnamefont {J.~S.}\ \bibnamefont
  {Moon}}, \bibinfo {author} {\bibfnamefont {L.}~\bibnamefont {Manuel}},
  \bibinfo {author} {\bibfnamefont {M.~J.}\ \bibnamefont {Churchfield}},
  \bibinfo {author} {\bibfnamefont {S.}~\bibnamefont {Lee}},\ and\ \bibinfo
  {author} {\bibfnamefont {P.~S.}\ \bibnamefont {Veers}},\ }\bibfield  {title}
  {\bibinfo {title} {Toward development of a stochastic wake model: Validation
  using les and turbine loads},\ }\href@noop {} {\bibfield  {journal} {\bibinfo
   {journal} {Energies}\ }\textbf {\bibinfo {volume} {11}},\ \bibinfo {pages}
  {53} (\bibinfo {year} {2017})}\BibitemShut {NoStop}%
\bibitem [{\citenamefont {Trefethen}\ \emph {et~al.}(1993)\citenamefont
  {Trefethen}, \citenamefont {Trefethen}, \citenamefont {Reddy},\ and\
  \citenamefont {Driscoll}}]{trefethen1993hydrodynamic}%
  \BibitemOpen
  \bibfield  {author} {\bibinfo {author} {\bibfnamefont {L.~N.}\ \bibnamefont
  {Trefethen}}, \bibinfo {author} {\bibfnamefont {A.~E.}\ \bibnamefont
  {Trefethen}}, \bibinfo {author} {\bibfnamefont {S.~C.}\ \bibnamefont
  {Reddy}},\ and\ \bibinfo {author} {\bibfnamefont {T.~A.}\ \bibnamefont
  {Driscoll}},\ }\bibfield  {title} {\bibinfo {title} {Hydrodynamic stability
  without eigenvalues},\ }\href@noop {} {\bibfield  {journal} {\bibinfo
  {journal} {Science}\ }\textbf {\bibinfo {volume} {261}},\ \bibinfo {pages}
  {578} (\bibinfo {year} {1993})}\BibitemShut {NoStop}%
\bibitem [{\citenamefont {Jovanovi{\'c}}\ and\ \citenamefont
  {Bamieh}(2005)}]{jovanovic2005componentwise}%
  \BibitemOpen
  \bibfield  {author} {\bibinfo {author} {\bibfnamefont {M.~R.}\ \bibnamefont
  {Jovanovi{\'c}}}\ and\ \bibinfo {author} {\bibfnamefont {B.}~\bibnamefont
  {Bamieh}},\ }\bibfield  {title} {\bibinfo {title} {Componentwise energy
  amplification in channel flows},\ }\href@noop {} {\bibfield  {journal}
  {\bibinfo  {journal} {Journal of Fluid Mechanics}\ }\textbf {\bibinfo
  {volume} {534}},\ \bibinfo {pages} {145} (\bibinfo {year}
  {2005})}\BibitemShut {NoStop}%
\bibitem [{\citenamefont {Schmid}\ \emph {et~al.}(2002)\citenamefont {Schmid},
  \citenamefont {Henningson},\ and\ \citenamefont
  {Jankowski}}]{schmid2002stability}%
  \BibitemOpen
  \bibfield  {author} {\bibinfo {author} {\bibfnamefont {P.~J.}\ \bibnamefont
  {Schmid}}, \bibinfo {author} {\bibfnamefont {D.~S.}\ \bibnamefont
  {Henningson}},\ and\ \bibinfo {author} {\bibfnamefont {D.}~\bibnamefont
  {Jankowski}},\ }\bibfield  {title} {\bibinfo {title} {Stability and
  transition in shear flows. applied mathematical sciences, vol. 142},\
  }\href@noop {} {\bibfield  {journal} {\bibinfo  {journal} {Appl. Mech. Rev.}\
  }\textbf {\bibinfo {volume} {55}},\ \bibinfo {pages} {B57} (\bibinfo {year}
  {2002})}\BibitemShut {NoStop}%
\bibitem [{\citenamefont {McKeon}\ and\ \citenamefont
  {Sharma}(2010)}]{mckeon2010critical}%
  \BibitemOpen
  \bibfield  {author} {\bibinfo {author} {\bibfnamefont {B.~J.}\ \bibnamefont
  {McKeon}}\ and\ \bibinfo {author} {\bibfnamefont {A.~S.}\ \bibnamefont
  {Sharma}},\ }\bibfield  {title} {\bibinfo {title} {A critical-layer framework
  for turbulent pipe flow},\ }\href@noop {} {\bibfield  {journal} {\bibinfo
  {journal} {Journal of Fluid Mechanics}\ }\textbf {\bibinfo {volume} {658}},\
  \bibinfo {pages} {336} (\bibinfo {year} {2010})}\BibitemShut {NoStop}%
\bibitem [{\citenamefont {Farrell}\ and\ \citenamefont
  {Ioannou}(1993)}]{farrell1993stochastic}%
  \BibitemOpen
  \bibfield  {author} {\bibinfo {author} {\bibfnamefont {B.~F.}\ \bibnamefont
  {Farrell}}\ and\ \bibinfo {author} {\bibfnamefont {P.~J.}\ \bibnamefont
  {Ioannou}},\ }\bibfield  {title} {\bibinfo {title} {Stochastic forcing of the
  linearized navier--stokes equations},\ }\href@noop {} {\bibfield  {journal}
  {\bibinfo  {journal} {Physics of Fluids A: Fluid Dynamics}\ }\textbf
  {\bibinfo {volume} {5}},\ \bibinfo {pages} {2600} (\bibinfo {year}
  {1993})}\BibitemShut {NoStop}%
\bibitem [{\citenamefont {Taira}\ \emph {et~al.}(2017)\citenamefont {Taira},
  \citenamefont {Brunton}, \citenamefont {Dawson}, \citenamefont {Rowley},
  \citenamefont {Colonius}, \citenamefont {McKeon}, \citenamefont {Schmidt},
  \citenamefont {Gordeyev}, \citenamefont {Theofilis},\ and\ \citenamefont
  {Ukeiley}}]{taira2017modal}%
  \BibitemOpen
  \bibfield  {author} {\bibinfo {author} {\bibfnamefont {K.}~\bibnamefont
  {Taira}}, \bibinfo {author} {\bibfnamefont {S.~L.}\ \bibnamefont {Brunton}},
  \bibinfo {author} {\bibfnamefont {S.~T.}\ \bibnamefont {Dawson}}, \bibinfo
  {author} {\bibfnamefont {C.~W.}\ \bibnamefont {Rowley}}, \bibinfo {author}
  {\bibfnamefont {T.}~\bibnamefont {Colonius}}, \bibinfo {author}
  {\bibfnamefont {B.~J.}\ \bibnamefont {McKeon}}, \bibinfo {author}
  {\bibfnamefont {O.~T.}\ \bibnamefont {Schmidt}}, \bibinfo {author}
  {\bibfnamefont {S.}~\bibnamefont {Gordeyev}}, \bibinfo {author}
  {\bibfnamefont {V.}~\bibnamefont {Theofilis}},\ and\ \bibinfo {author}
  {\bibfnamefont {L.~S.}\ \bibnamefont {Ukeiley}},\ }\bibfield  {title}
  {\bibinfo {title} {Modal analysis of fluid flows: An overview},\ }\href@noop
  {} {\bibfield  {journal} {\bibinfo  {journal} {Aiaa Journal}\ }\textbf
  {\bibinfo {volume} {55}},\ \bibinfo {pages} {4013} (\bibinfo {year}
  {2017})}\BibitemShut {NoStop}%
\bibitem [{\citenamefont {Hwang}\ and\ \citenamefont
  {Cossu}(2010{\natexlab{a}})}]{hwang2010linear}%
  \BibitemOpen
  \bibfield  {author} {\bibinfo {author} {\bibfnamefont {Y.}~\bibnamefont
  {Hwang}}\ and\ \bibinfo {author} {\bibfnamefont {C.}~\bibnamefont {Cossu}},\
  }\bibfield  {title} {\bibinfo {title} {Linear non-normal energy amplification
  of harmonic and stochastic forcing in the turbulent channel flow},\
  }\href@noop {} {\bibfield  {journal} {\bibinfo  {journal} {Journal of Fluid
  Mechanics}\ }\textbf {\bibinfo {volume} {664}},\ \bibinfo {pages} {51}
  (\bibinfo {year} {2010}{\natexlab{a}})}\BibitemShut {NoStop}%
\bibitem [{\citenamefont {Hwang}\ and\ \citenamefont
  {Cossu}(2010{\natexlab{b}})}]{hwang2010amplification}%
  \BibitemOpen
  \bibfield  {author} {\bibinfo {author} {\bibfnamefont {Y.}~\bibnamefont
  {Hwang}}\ and\ \bibinfo {author} {\bibfnamefont {C.}~\bibnamefont {Cossu}},\
  }\bibfield  {title} {\bibinfo {title} {Amplification of coherent streaks in
  the turbulent couette flow: an input--output analysis at low reynolds
  number},\ }\href@noop {} {\bibfield  {journal} {\bibinfo  {journal} {Journal
  of Fluid Mechanics}\ }\textbf {\bibinfo {volume} {643}},\ \bibinfo {pages}
  {333} (\bibinfo {year} {2010}{\natexlab{b}})}\BibitemShut {NoStop}%
\bibitem [{\citenamefont {Sharma}\ and\ \citenamefont
  {McKeon}(2013)}]{sharma2013coherent}%
  \BibitemOpen
  \bibfield  {author} {\bibinfo {author} {\bibfnamefont {A.}~\bibnamefont
  {Sharma}}\ and\ \bibinfo {author} {\bibfnamefont {B.~J.}\ \bibnamefont
  {McKeon}},\ }\bibfield  {title} {\bibinfo {title} {On coherent structure in
  wall turbulence},\ }\href@noop {} {\bibfield  {journal} {\bibinfo  {journal}
  {Journal of Fluid Mechanics}\ }\textbf {\bibinfo {volume} {728}},\ \bibinfo
  {pages} {196} (\bibinfo {year} {2013})}\BibitemShut {NoStop}%
\bibitem [{\citenamefont {Jeun}\ \emph {et~al.}(2016)\citenamefont {Jeun},
  \citenamefont {Nichols},\ and\ \citenamefont
  {Jovanovi{\'c}}}]{jeun2016input}%
  \BibitemOpen
  \bibfield  {author} {\bibinfo {author} {\bibfnamefont {J.}~\bibnamefont
  {Jeun}}, \bibinfo {author} {\bibfnamefont {J.~W.}\ \bibnamefont {Nichols}},\
  and\ \bibinfo {author} {\bibfnamefont {M.~R.}\ \bibnamefont
  {Jovanovi{\'c}}},\ }\bibfield  {title} {\bibinfo {title} {Input-output
  analysis of high-speed axisymmetric isothermal jet noise},\ }\href@noop {}
  {\bibfield  {journal} {\bibinfo  {journal} {Physics of Fluids}\ }\textbf
  {\bibinfo {volume} {28}},\ \bibinfo {pages} {047101} (\bibinfo {year}
  {2016})}\BibitemShut {NoStop}%
\bibitem [{\citenamefont {Schmidt}\ \emph {et~al.}(2018)\citenamefont
  {Schmidt}, \citenamefont {Towne}, \citenamefont {Rigas}, \citenamefont
  {Colonius},\ and\ \citenamefont {Br{\`e}s}}]{schmidt2018spectral}%
  \BibitemOpen
  \bibfield  {author} {\bibinfo {author} {\bibfnamefont {O.~T.}\ \bibnamefont
  {Schmidt}}, \bibinfo {author} {\bibfnamefont {A.}~\bibnamefont {Towne}},
  \bibinfo {author} {\bibfnamefont {G.}~\bibnamefont {Rigas}}, \bibinfo
  {author} {\bibfnamefont {T.}~\bibnamefont {Colonius}},\ and\ \bibinfo
  {author} {\bibfnamefont {G.~A.}\ \bibnamefont {Br{\`e}s}},\ }\bibfield
  {title} {\bibinfo {title} {Spectral analysis of jet turbulence},\ }\href@noop
  {} {\bibfield  {journal} {\bibinfo  {journal} {Journal of Fluid Mechanics}\
  }\textbf {\bibinfo {volume} {855}},\ \bibinfo {pages} {953} (\bibinfo {year}
  {2018})}\BibitemShut {NoStop}%
\bibitem [{\citenamefont {Garnaud}\ \emph {et~al.}(2013)\citenamefont
  {Garnaud}, \citenamefont {Lesshafft}, \citenamefont {Schmid},\ and\
  \citenamefont {Huerre}}]{garnaud2013preferred}%
  \BibitemOpen
  \bibfield  {author} {\bibinfo {author} {\bibfnamefont {X.}~\bibnamefont
  {Garnaud}}, \bibinfo {author} {\bibfnamefont {L.}~\bibnamefont {Lesshafft}},
  \bibinfo {author} {\bibfnamefont {P.}~\bibnamefont {Schmid}},\ and\ \bibinfo
  {author} {\bibfnamefont {P.}~\bibnamefont {Huerre}},\ }\bibfield  {title}
  {\bibinfo {title} {The preferred mode of incompressible jets: linear
  frequency response analysis},\ }\href@noop {} {\bibfield  {journal} {\bibinfo
   {journal} {Journal of Fluid Mechanics}\ }\textbf {\bibinfo {volume} {716}},\
  \bibinfo {pages} {189} (\bibinfo {year} {2013})}\BibitemShut {NoStop}%
\bibitem [{\citenamefont {Jin}\ \emph {et~al.}(2021)\citenamefont {Jin},
  \citenamefont {Symon},\ and\ \citenamefont {Illingworth}}]{jin2021energy}%
  \BibitemOpen
  \bibfield  {author} {\bibinfo {author} {\bibfnamefont {B.}~\bibnamefont
  {Jin}}, \bibinfo {author} {\bibfnamefont {S.}~\bibnamefont {Symon}},\ and\
  \bibinfo {author} {\bibfnamefont {S.~J.}\ \bibnamefont {Illingworth}},\
  }\bibfield  {title} {\bibinfo {title} {Energy transfer mechanisms and
  resolvent analysis in the cylinder wake},\ }\href@noop {} {\bibfield
  {journal} {\bibinfo  {journal} {Physical Review Fluids}\ }\textbf {\bibinfo
  {volume} {6}},\ \bibinfo {pages} {024702} (\bibinfo {year}
  {2021})}\BibitemShut {NoStop}%
\bibitem [{\citenamefont {Symon}\ \emph {et~al.}(2018)\citenamefont {Symon},
  \citenamefont {Rosenberg}, \citenamefont {Dawson},\ and\ \citenamefont
  {McKeon}}]{symon2018non}%
  \BibitemOpen
  \bibfield  {author} {\bibinfo {author} {\bibfnamefont {S.}~\bibnamefont
  {Symon}}, \bibinfo {author} {\bibfnamefont {K.}~\bibnamefont {Rosenberg}},
  \bibinfo {author} {\bibfnamefont {S.~T.}\ \bibnamefont {Dawson}},\ and\
  \bibinfo {author} {\bibfnamefont {B.~J.}\ \bibnamefont {McKeon}},\ }\bibfield
   {title} {\bibinfo {title} {Non-normality and classification of amplification
  mechanisms in stability and resolvent analysis},\ }\href@noop {} {\bibfield
  {journal} {\bibinfo  {journal} {Physical Review Fluids}\ }\textbf {\bibinfo
  {volume} {3}},\ \bibinfo {pages} {053902} (\bibinfo {year}
  {2018})}\BibitemShut {NoStop}%
\bibitem [{\citenamefont {De~Cillis}\ \emph {et~al.}(2022)\citenamefont
  {De~Cillis}, \citenamefont {Cherubini}, \citenamefont {Semeraro},
  \citenamefont {Leonardi},\ and\ \citenamefont {De~Palma}}]{de2022stability}%
  \BibitemOpen
  \bibfield  {author} {\bibinfo {author} {\bibfnamefont {G.}~\bibnamefont
  {De~Cillis}}, \bibinfo {author} {\bibfnamefont {S.}~\bibnamefont
  {Cherubini}}, \bibinfo {author} {\bibfnamefont {O.}~\bibnamefont {Semeraro}},
  \bibinfo {author} {\bibfnamefont {S.}~\bibnamefont {Leonardi}},\ and\
  \bibinfo {author} {\bibfnamefont {P.}~\bibnamefont {De~Palma}},\ }\bibfield
  {title} {\bibinfo {title} {Stability and optimal forcing analysis of a wind
  turbine wake: Comparison with pod},\ }\href@noop {} {\bibfield  {journal}
  {\bibinfo  {journal} {Renewable Energy}\ }\textbf {\bibinfo {volume} {181}},\
  \bibinfo {pages} {765} (\bibinfo {year} {2022})}\BibitemShut {NoStop}%
\bibitem [{\citenamefont {G{\'o}mez}\ \emph {et~al.}(2016)\citenamefont
  {G{\'o}mez}, \citenamefont {Blackburn}, \citenamefont {Rudman}, \citenamefont
  {Sharma},\ and\ \citenamefont {McKeon}}]{gomez2016reduced}%
  \BibitemOpen
  \bibfield  {author} {\bibinfo {author} {\bibfnamefont {F.}~\bibnamefont
  {G{\'o}mez}}, \bibinfo {author} {\bibfnamefont {H.}~\bibnamefont
  {Blackburn}}, \bibinfo {author} {\bibfnamefont {M.}~\bibnamefont {Rudman}},
  \bibinfo {author} {\bibfnamefont {A.}~\bibnamefont {Sharma}},\ and\ \bibinfo
  {author} {\bibfnamefont {B.}~\bibnamefont {McKeon}},\ }\bibfield  {title}
  {\bibinfo {title} {A reduced-order model of three-dimensional unsteady flow
  in a cavity based on the resolvent operator},\ }\href@noop {} {\bibfield
  {journal} {\bibinfo  {journal} {Journal of Fluid Mechanics}\ }\textbf
  {\bibinfo {volume} {798}},\ \bibinfo {pages} {R2} (\bibinfo {year}
  {2016})}\BibitemShut {NoStop}%
\bibitem [{\citenamefont {Jovanovi{\'c}}(2021)}]{jovanovic2021bypass}%
  \BibitemOpen
  \bibfield  {author} {\bibinfo {author} {\bibfnamefont {M.~R.}\ \bibnamefont
  {Jovanovi{\'c}}},\ }\bibfield  {title} {\bibinfo {title} {From bypass
  transition to flow control and data-driven turbulence modeling: an
  input--output viewpoint},\ }\href@noop {} {\bibfield  {journal} {\bibinfo
  {journal} {Annual Review of Fluid Mechanics}\ }\textbf {\bibinfo {volume}
  {53}},\ \bibinfo {pages} {311} (\bibinfo {year} {2021})}\BibitemShut
  {NoStop}%
\bibitem [{\citenamefont {Moarref}\ \emph {et~al.}(2013)\citenamefont
  {Moarref}, \citenamefont {Sharma}, \citenamefont {Tropp},\ and\ \citenamefont
  {McKeon}}]{moarref2013model}%
  \BibitemOpen
  \bibfield  {author} {\bibinfo {author} {\bibfnamefont {R.}~\bibnamefont
  {Moarref}}, \bibinfo {author} {\bibfnamefont {A.~S.}\ \bibnamefont {Sharma}},
  \bibinfo {author} {\bibfnamefont {J.~A.}\ \bibnamefont {Tropp}},\ and\
  \bibinfo {author} {\bibfnamefont {B.~J.}\ \bibnamefont {McKeon}},\ }\bibfield
   {title} {\bibinfo {title} {Model-based scaling of the streamwise energy
  density in high-reynolds-number turbulent channels},\ }\href@noop {}
  {\bibfield  {journal} {\bibinfo  {journal} {Journal of Fluid Mechanics}\
  }\textbf {\bibinfo {volume} {734}},\ \bibinfo {pages} {275} (\bibinfo {year}
  {2013})}\BibitemShut {NoStop}%
\bibitem [{\citenamefont {Pickering}\ \emph
  {et~al.}(2021{\natexlab{a}})\citenamefont {Pickering}, \citenamefont {Towne},
  \citenamefont {Jordan},\ and\ \citenamefont
  {Colonius}}]{pickering2021resolvent}%
  \BibitemOpen
  \bibfield  {author} {\bibinfo {author} {\bibfnamefont {E.}~\bibnamefont
  {Pickering}}, \bibinfo {author} {\bibfnamefont {A.}~\bibnamefont {Towne}},
  \bibinfo {author} {\bibfnamefont {P.}~\bibnamefont {Jordan}},\ and\ \bibinfo
  {author} {\bibfnamefont {T.}~\bibnamefont {Colonius}},\ }\bibfield  {title}
  {\bibinfo {title} {Resolvent-based modeling of turbulent jet noise},\
  }\href@noop {} {\bibfield  {journal} {\bibinfo  {journal} {The Journal of the
  Acoustical Society of America}\ }\textbf {\bibinfo {volume} {150}},\ \bibinfo
  {pages} {2421} (\bibinfo {year} {2021}{\natexlab{a}})}\BibitemShut {NoStop}%
\bibitem [{\citenamefont {Pickering}\ \emph
  {et~al.}(2021{\natexlab{b}})\citenamefont {Pickering}, \citenamefont {Rigas},
  \citenamefont {Schmidt}, \citenamefont {Sipp},\ and\ \citenamefont
  {Colonius}}]{pickering2021optimal}%
  \BibitemOpen
  \bibfield  {author} {\bibinfo {author} {\bibfnamefont {E.}~\bibnamefont
  {Pickering}}, \bibinfo {author} {\bibfnamefont {G.}~\bibnamefont {Rigas}},
  \bibinfo {author} {\bibfnamefont {O.~T.}\ \bibnamefont {Schmidt}}, \bibinfo
  {author} {\bibfnamefont {D.}~\bibnamefont {Sipp}},\ and\ \bibinfo {author}
  {\bibfnamefont {T.}~\bibnamefont {Colonius}},\ }\bibfield  {title} {\bibinfo
  {title} {Optimal eddy viscosity for resolvent-based models of coherent
  structures in turbulent jets},\ }\href@noop {} {\bibfield  {journal}
  {\bibinfo  {journal} {Journal of Fluid Mechanics}\ }\textbf {\bibinfo
  {volume} {917}},\ \bibinfo {pages} {A29} (\bibinfo {year}
  {2021}{\natexlab{b}})}\BibitemShut {NoStop}%
\bibitem [{\citenamefont {Towne}\ \emph {et~al.}(2018)\citenamefont {Towne},
  \citenamefont {Schmidt},\ and\ \citenamefont {Colonius}}]{towne2018spectral}%
  \BibitemOpen
  \bibfield  {author} {\bibinfo {author} {\bibfnamefont {A.}~\bibnamefont
  {Towne}}, \bibinfo {author} {\bibfnamefont {O.~T.}\ \bibnamefont {Schmidt}},\
  and\ \bibinfo {author} {\bibfnamefont {T.}~\bibnamefont {Colonius}},\
  }\bibfield  {title} {\bibinfo {title} {Spectral proper orthogonal
  decomposition and its relationship to dynamic mode decomposition and
  resolvent analysis},\ }\href@noop {} {\bibfield  {journal} {\bibinfo
  {journal} {Journal of Fluid Mechanics}\ }\textbf {\bibinfo {volume} {847}},\
  \bibinfo {pages} {821} (\bibinfo {year} {2018})}\BibitemShut {NoStop}%
\bibitem [{\citenamefont {McKeon}(2017)}]{mckeon2017engine}%
  \BibitemOpen
  \bibfield  {author} {\bibinfo {author} {\bibfnamefont {B.}~\bibnamefont
  {McKeon}},\ }\bibfield  {title} {\bibinfo {title} {The engine behind (wall)
  turbulence: perspectives on scale interactions},\ }\href@noop {} {\bibfield
  {journal} {\bibinfo  {journal} {Journal of Fluid Mechanics}\ }\textbf
  {\bibinfo {volume} {817}},\ \bibinfo {pages} {P1} (\bibinfo {year}
  {2017})}\BibitemShut {NoStop}%
\bibitem [{\citenamefont {Towne}\ \emph {et~al.}(2017)\citenamefont {Towne},
  \citenamefont {Bres},\ and\ \citenamefont {Lele}}]{towne2017statistical}%
  \BibitemOpen
  \bibfield  {author} {\bibinfo {author} {\bibfnamefont {A.}~\bibnamefont
  {Towne}}, \bibinfo {author} {\bibfnamefont {G.~A.}\ \bibnamefont {Bres}},\
  and\ \bibinfo {author} {\bibfnamefont {S.~K.}\ \bibnamefont {Lele}},\
  }\bibfield  {title} {\bibinfo {title} {A statistical jet-noise model based on
  the resolvent framework},\ }in\ \href@noop {} {\emph {\bibinfo {booktitle}
  {23rd AIAA/CEAS Aeroacoustics Conference}}}\ (\bibinfo {year} {2017})\ p.\
  \bibinfo {pages} {3706}\BibitemShut {NoStop}%
\bibitem [{\citenamefont {Zare}\ \emph {et~al.}(2017)\citenamefont {Zare},
  \citenamefont {Jovanovi{\'c}},\ and\ \citenamefont
  {Georgiou}}]{zare2017colour}%
  \BibitemOpen
  \bibfield  {author} {\bibinfo {author} {\bibfnamefont {A.}~\bibnamefont
  {Zare}}, \bibinfo {author} {\bibfnamefont {M.~R.}\ \bibnamefont
  {Jovanovi{\'c}}},\ and\ \bibinfo {author} {\bibfnamefont {T.~T.}\
  \bibnamefont {Georgiou}},\ }\bibfield  {title} {\bibinfo {title} {Colour of
  turbulence},\ }\href@noop {} {\bibfield  {journal} {\bibinfo  {journal}
  {Journal of Fluid Mechanics}\ }\textbf {\bibinfo {volume} {812}},\ \bibinfo
  {pages} {636} (\bibinfo {year} {2017})}\BibitemShut {NoStop}%
\bibitem [{\citenamefont {Towne}\ \emph {et~al.}(2020)\citenamefont {Towne},
  \citenamefont {Lozano-Dur{\'a}n},\ and\ \citenamefont
  {Yang}}]{towne2020resolvent}%
  \BibitemOpen
  \bibfield  {author} {\bibinfo {author} {\bibfnamefont {A.}~\bibnamefont
  {Towne}}, \bibinfo {author} {\bibfnamefont {A.}~\bibnamefont
  {Lozano-Dur{\'a}n}},\ and\ \bibinfo {author} {\bibfnamefont {X.}~\bibnamefont
  {Yang}},\ }\bibfield  {title} {\bibinfo {title} {Resolvent-based estimation
  of space--time flow statistics},\ }\href@noop {} {\bibfield  {journal}
  {\bibinfo  {journal} {Journal of Fluid Mechanics}\ }\textbf {\bibinfo
  {volume} {883}},\ \bibinfo {pages} {A17} (\bibinfo {year}
  {2020})}\BibitemShut {NoStop}%
\bibitem [{\citenamefont {Moarref}\ and\ \citenamefont
  {Jovanovi{\'c}}(2012)}]{moarref2012model}%
  \BibitemOpen
  \bibfield  {author} {\bibinfo {author} {\bibfnamefont {R.}~\bibnamefont
  {Moarref}}\ and\ \bibinfo {author} {\bibfnamefont {M.~R.}\ \bibnamefont
  {Jovanovi{\'c}}},\ }\bibfield  {title} {\bibinfo {title} {Model-based design
  of transverse wall oscillations for turbulent drag reduction},\ }\href@noop
  {} {\bibfield  {journal} {\bibinfo  {journal} {Journal of fluid mechanics}\
  }\textbf {\bibinfo {volume} {707}},\ \bibinfo {pages} {205} (\bibinfo {year}
  {2012})}\BibitemShut {NoStop}%
\bibitem [{\citenamefont {Yim}\ \emph {et~al.}(2019)\citenamefont {Yim},
  \citenamefont {Meliga},\ and\ \citenamefont {Gallaire}}]{yim2019self}%
  \BibitemOpen
  \bibfield  {author} {\bibinfo {author} {\bibfnamefont {E.}~\bibnamefont
  {Yim}}, \bibinfo {author} {\bibfnamefont {P.}~\bibnamefont {Meliga}},\ and\
  \bibinfo {author} {\bibfnamefont {F.}~\bibnamefont {Gallaire}},\ }\bibfield
  {title} {\bibinfo {title} {Self-consistent triple decomposition of the
  turbulent flow over a backward-facing step under finite amplitude harmonic
  forcing},\ }\href@noop {} {\bibfield  {journal} {\bibinfo  {journal}
  {Proceedings of the Royal Society A}\ }\textbf {\bibinfo {volume} {475}},\
  \bibinfo {pages} {20190018} (\bibinfo {year} {2019})}\BibitemShut {NoStop}%
\bibitem [{\citenamefont {Morra}\ \emph {et~al.}(2019)\citenamefont {Morra},
  \citenamefont {Semeraro}, \citenamefont {Henningson},\ and\ \citenamefont
  {Cossu}}]{morra2019relevance}%
  \BibitemOpen
  \bibfield  {author} {\bibinfo {author} {\bibfnamefont {P.}~\bibnamefont
  {Morra}}, \bibinfo {author} {\bibfnamefont {O.}~\bibnamefont {Semeraro}},
  \bibinfo {author} {\bibfnamefont {D.~S.}\ \bibnamefont {Henningson}},\ and\
  \bibinfo {author} {\bibfnamefont {C.}~\bibnamefont {Cossu}},\ }\bibfield
  {title} {\bibinfo {title} {On the relevance of reynolds stresses in resolvent
  analyses of turbulent wall-bounded flows},\ }\href@noop {} {\bibfield
  {journal} {\bibinfo  {journal} {Journal of Fluid Mechanics}\ }\textbf
  {\bibinfo {volume} {867}},\ \bibinfo {pages} {969} (\bibinfo {year}
  {2019})}\BibitemShut {NoStop}%
\bibitem [{\citenamefont {Reynolds}\ and\ \citenamefont
  {Hussain}(1972)}]{reynolds1972mechanics}%
  \BibitemOpen
  \bibfield  {author} {\bibinfo {author} {\bibfnamefont {W.}~\bibnamefont
  {Reynolds}}\ and\ \bibinfo {author} {\bibfnamefont {A.}~\bibnamefont
  {Hussain}},\ }\bibfield  {title} {\bibinfo {title} {The mechanics of an
  organized wave in turbulent shear flow. part 3. theoretical models and
  comparisons with experiments},\ }\href@noop {} {\bibfield  {journal}
  {\bibinfo  {journal} {Journal of Fluid Mechanics}\ }\textbf {\bibinfo
  {volume} {54}},\ \bibinfo {pages} {263} (\bibinfo {year} {1972})}\BibitemShut
  {NoStop}%
\bibitem [{\citenamefont {Reynolds}\ and\ \citenamefont
  {Tiederman}(1967)}]{reynolds1967stability}%
  \BibitemOpen
  \bibfield  {author} {\bibinfo {author} {\bibfnamefont {W.}~\bibnamefont
  {Reynolds}}\ and\ \bibinfo {author} {\bibfnamefont {W.}~\bibnamefont
  {Tiederman}},\ }\bibfield  {title} {\bibinfo {title} {Stability of turbulent
  channel flow, with application to malkus's theory},\ }\href@noop {}
  {\bibfield  {journal} {\bibinfo  {journal} {Journal of Fluid Mechanics}\
  }\textbf {\bibinfo {volume} {27}},\ \bibinfo {pages} {253} (\bibinfo {year}
  {1967})}\BibitemShut {NoStop}%
\bibitem [{\citenamefont {Jensen}(1983)}]{jensen1983note}%
  \BibitemOpen
  \bibfield  {author} {\bibinfo {author} {\bibfnamefont {N.~O.}\ \bibnamefont
  {Jensen}},\ }\href@noop {} {\emph {\bibinfo {title} {A note on wind generator
  interaction}}},\ Vol.\ \bibinfo {volume} {2411}\ (\bibinfo  {publisher}
  {Citeseer},\ \bibinfo {year} {1983})\BibitemShut {NoStop}%
\bibitem [{\citenamefont {Bastankhah}\ and\ \citenamefont
  {Port{\'e}-Agel}(2014)}]{bastankhah2014new}%
  \BibitemOpen
  \bibfield  {author} {\bibinfo {author} {\bibfnamefont {M.}~\bibnamefont
  {Bastankhah}}\ and\ \bibinfo {author} {\bibfnamefont {F.}~\bibnamefont
  {Port{\'e}-Agel}},\ }\bibfield  {title} {\bibinfo {title} {A new analytical
  model for wind-turbine wakes},\ }\href@noop {} {\bibfield  {journal}
  {\bibinfo  {journal} {Renewable energy}\ }\textbf {\bibinfo {volume} {70}},\
  \bibinfo {pages} {116} (\bibinfo {year} {2014})}\BibitemShut {NoStop}%
\bibitem [{\citenamefont {Larsen}\ \emph {et~al.}(2007)\citenamefont {Larsen},
  \citenamefont {Aagaard~Madsen},\ and\ \citenamefont
  {Bing{\"o}l}}]{larsen2007dynamic}%
  \BibitemOpen
  \bibfield  {author} {\bibinfo {author} {\bibfnamefont {G.~C.}\ \bibnamefont
  {Larsen}}, \bibinfo {author} {\bibfnamefont {H.}~\bibnamefont
  {Aagaard~Madsen}},\ and\ \bibinfo {author} {\bibfnamefont {F.}~\bibnamefont
  {Bing{\"o}l}},\ }\bibfield  {title} {\bibinfo {title} {Dynamic wake
  meandering modeling},\ }\href@noop {} {\  (\bibinfo {year}
  {2007})}\BibitemShut {NoStop}%
\bibitem [{\citenamefont {Jonkman}\ \emph {et~al.}(2017)\citenamefont
  {Jonkman}, \citenamefont {Annoni}, \citenamefont {Hayman}, \citenamefont
  {Jonkman},\ and\ \citenamefont {Purkayastha}}]{jonkman2017development}%
  \BibitemOpen
  \bibfield  {author} {\bibinfo {author} {\bibfnamefont {J.~M.}\ \bibnamefont
  {Jonkman}}, \bibinfo {author} {\bibfnamefont {J.}~\bibnamefont {Annoni}},
  \bibinfo {author} {\bibfnamefont {G.}~\bibnamefont {Hayman}}, \bibinfo
  {author} {\bibfnamefont {B.}~\bibnamefont {Jonkman}},\ and\ \bibinfo {author}
  {\bibfnamefont {A.}~\bibnamefont {Purkayastha}},\ }\bibfield  {title}
  {\bibinfo {title} {Development of fast. farm: A new multi-physics engineering
  tool for wind-farm design and analysis},\ }in\ \href@noop {} {\emph {\bibinfo
  {booktitle} {35th wind energy symposium}}}\ (\bibinfo {year} {2017})\ p.\
  \bibinfo {pages} {0454}\BibitemShut {NoStop}%
\bibitem [{\citenamefont {Shaler}\ and\ \citenamefont
  {Jonkman}(2021)}]{shaler2021fast}%
  \BibitemOpen
  \bibfield  {author} {\bibinfo {author} {\bibfnamefont {K.}~\bibnamefont
  {Shaler}}\ and\ \bibinfo {author} {\bibfnamefont {J.}~\bibnamefont
  {Jonkman}},\ }\bibfield  {title} {\bibinfo {title} {Fast. farm development
  and validation of structural load prediction against large eddy
  simulations},\ }\href@noop {} {\bibfield  {journal} {\bibinfo  {journal}
  {Wind Energy}\ }\textbf {\bibinfo {volume} {24}},\ \bibinfo {pages} {428}
  (\bibinfo {year} {2021})}\BibitemShut {NoStop}%
\bibitem [{\citenamefont {Churchfield}\ \emph {et~al.}(2012)\citenamefont
  {Churchfield}, \citenamefont {Lee}, \citenamefont {Michalakes},\ and\
  \citenamefont {Moriarty}}]{churchfield2012numerical}%
  \BibitemOpen
  \bibfield  {author} {\bibinfo {author} {\bibfnamefont {M.~J.}\ \bibnamefont
  {Churchfield}}, \bibinfo {author} {\bibfnamefont {S.}~\bibnamefont {Lee}},
  \bibinfo {author} {\bibfnamefont {J.}~\bibnamefont {Michalakes}},\ and\
  \bibinfo {author} {\bibfnamefont {P.~J.}\ \bibnamefont {Moriarty}},\
  }\bibfield  {title} {\bibinfo {title} {A numerical study of the effects of
  atmospheric and wake turbulence on wind turbine dynamics},\ }\href@noop {}
  {\bibfield  {journal} {\bibinfo  {journal} {Journal of turbulence}\ ,\
  \bibinfo {pages} {N14}} (\bibinfo {year} {2012})}\BibitemShut {NoStop}%
\bibitem [{\citenamefont {Jasak}(1996)}]{jasak1996error}%
  \BibitemOpen
  \bibfield  {author} {\bibinfo {author} {\bibfnamefont {H.}~\bibnamefont
  {Jasak}},\ }\bibfield  {title} {\bibinfo {title} {Error analysis and
  estimation for the finite volume method with applications to fluid flows.},\
  }\href@noop {} {\  (\bibinfo {year} {1996})}\BibitemShut {NoStop}%
\bibitem [{\citenamefont {Bou-Zeid}\ \emph {et~al.}(2005)\citenamefont
  {Bou-Zeid}, \citenamefont {Meneveau},\ and\ \citenamefont
  {Parlange}}]{bou2005scale}%
  \BibitemOpen
  \bibfield  {author} {\bibinfo {author} {\bibfnamefont {E.}~\bibnamefont
  {Bou-Zeid}}, \bibinfo {author} {\bibfnamefont {C.}~\bibnamefont {Meneveau}},\
  and\ \bibinfo {author} {\bibfnamefont {M.}~\bibnamefont {Parlange}},\
  }\bibfield  {title} {\bibinfo {title} {A scale-dependent lagrangian dynamic
  model for large eddy simulation of complex turbulent flows},\ }\href@noop {}
  {\bibfield  {journal} {\bibinfo  {journal} {Physics of fluids}\ }\textbf
  {\bibinfo {volume} {17}},\ \bibinfo {pages} {025105} (\bibinfo {year}
  {2005})}\BibitemShut {NoStop}%
\bibitem [{\citenamefont {Moeng}(1984)}]{moeng1984large}%
  \BibitemOpen
  \bibfield  {author} {\bibinfo {author} {\bibfnamefont {C.-H.}\ \bibnamefont
  {Moeng}},\ }\bibfield  {title} {\bibinfo {title} {A large-eddy-simulation
  model for the study of planetary boundary-layer turbulence},\ }\href@noop {}
  {\bibfield  {journal} {\bibinfo  {journal} {Journal of the Atmospheric
  Sciences}\ }\textbf {\bibinfo {volume} {41}},\ \bibinfo {pages} {2052}
  (\bibinfo {year} {1984})}\BibitemShut {NoStop}%
\bibitem [{\citenamefont {Kawai}\ and\ \citenamefont
  {Larsson}(2012)}]{kawai2012wall}%
  \BibitemOpen
  \bibfield  {author} {\bibinfo {author} {\bibfnamefont {S.}~\bibnamefont
  {Kawai}}\ and\ \bibinfo {author} {\bibfnamefont {J.}~\bibnamefont
  {Larsson}},\ }\bibfield  {title} {\bibinfo {title} {Wall-modeling in large
  eddy simulation: Length scales, grid resolution, and accuracy},\ }\href@noop
  {} {\bibfield  {journal} {\bibinfo  {journal} {Physics of Fluids}\ }\textbf
  {\bibinfo {volume} {24}},\ \bibinfo {pages} {015105} (\bibinfo {year}
  {2012})}\BibitemShut {NoStop}%
\bibitem [{\citenamefont {Wang}\ \emph {et~al.}(2023)\citenamefont {Wang},
  \citenamefont {Feng}, \citenamefont {Peng}, \citenamefont {Mao},
  \citenamefont {Doranehgard}, \citenamefont {Gupta}, \citenamefont {Li},\ and\
  \citenamefont {Wan}}]{wang2023implications}%
  \BibitemOpen
  \bibfield  {author} {\bibinfo {author} {\bibfnamefont {D.}~\bibnamefont
  {Wang}}, \bibinfo {author} {\bibfnamefont {D.}~\bibnamefont {Feng}}, \bibinfo
  {author} {\bibfnamefont {H.}~\bibnamefont {Peng}}, \bibinfo {author}
  {\bibfnamefont {F.}~\bibnamefont {Mao}}, \bibinfo {author} {\bibfnamefont
  {M.~H.}\ \bibnamefont {Doranehgard}}, \bibinfo {author} {\bibfnamefont
  {V.}~\bibnamefont {Gupta}}, \bibinfo {author} {\bibfnamefont {L.~K.}\
  \bibnamefont {Li}},\ and\ \bibinfo {author} {\bibfnamefont {M.}~\bibnamefont
  {Wan}},\ }\bibfield  {title} {\bibinfo {title} {Implications of steep hilly
  terrain for modeling wind-turbine wakes},\ }\href@noop {} {\bibfield
  {journal} {\bibinfo  {journal} {Journal of Cleaner Production}\ }\textbf
  {\bibinfo {volume} {398}},\ \bibinfo {pages} {136614} (\bibinfo {year}
  {2023})}\BibitemShut {NoStop}%
\bibitem [{\citenamefont {Jonkman}\ \emph {et~al.}(2009)\citenamefont
  {Jonkman}, \citenamefont {Butterfield}, \citenamefont {Musial},\ and\
  \citenamefont {Scott}}]{jonkman2009definition}%
  \BibitemOpen
  \bibfield  {author} {\bibinfo {author} {\bibfnamefont {J.}~\bibnamefont
  {Jonkman}}, \bibinfo {author} {\bibfnamefont {S.}~\bibnamefont
  {Butterfield}}, \bibinfo {author} {\bibfnamefont {W.}~\bibnamefont
  {Musial}},\ and\ \bibinfo {author} {\bibfnamefont {G.}~\bibnamefont
  {Scott}},\ }\href@noop {} {\emph {\bibinfo {title} {Definition of a 5-MW
  reference wind turbine for offshore system development}}},\ \bibinfo {type}
  {Tech. Rep.}\ (\bibinfo  {institution} {National Renewable Energy Lab.(NREL),
  Golden, CO (United States)},\ \bibinfo {year} {2009})\BibitemShut {NoStop}%
\bibitem [{\citenamefont {Wu}\ and\ \citenamefont
  {Port{\'e}-Agel}(2011)}]{wu2011large}%
  \BibitemOpen
  \bibfield  {author} {\bibinfo {author} {\bibfnamefont {Y.-T.}\ \bibnamefont
  {Wu}}\ and\ \bibinfo {author} {\bibfnamefont {F.}~\bibnamefont
  {Port{\'e}-Agel}},\ }\bibfield  {title} {\bibinfo {title} {Large-eddy
  simulation of wind-turbine wakes: evaluation of turbine parametrisations},\
  }\href@noop {} {\bibfield  {journal} {\bibinfo  {journal} {Boundary-layer
  meteorology}\ }\textbf {\bibinfo {volume} {138}},\ \bibinfo {pages} {345}
  (\bibinfo {year} {2011})}\BibitemShut {NoStop}%
\bibitem [{\citenamefont {Port{\'e}-Agel}\ \emph {et~al.}(2011)\citenamefont
  {Port{\'e}-Agel}, \citenamefont {Wu}, \citenamefont {Lu},\ and\ \citenamefont
  {Conzemius}}]{porte2011large}%
  \BibitemOpen
  \bibfield  {author} {\bibinfo {author} {\bibfnamefont {F.}~\bibnamefont
  {Port{\'e}-Agel}}, \bibinfo {author} {\bibfnamefont {Y.-T.}\ \bibnamefont
  {Wu}}, \bibinfo {author} {\bibfnamefont {H.}~\bibnamefont {Lu}},\ and\
  \bibinfo {author} {\bibfnamefont {R.~J.}\ \bibnamefont {Conzemius}},\
  }\bibfield  {title} {\bibinfo {title} {Large-eddy simulation of atmospheric
  boundary layer flow through wind turbines and wind farms},\ }\href@noop {}
  {\bibfield  {journal} {\bibinfo  {journal} {Journal of Wind Engineering and
  Industrial Aerodynamics}\ }\textbf {\bibinfo {volume} {99}},\ \bibinfo
  {pages} {154} (\bibinfo {year} {2011})}\BibitemShut {NoStop}%
\bibitem [{\citenamefont {Stevens}\ \emph {et~al.}(2018)\citenamefont
  {Stevens}, \citenamefont {Mart{\'\i}nez-Tossas},\ and\ \citenamefont
  {Meneveau}}]{stevens2018comparison}%
  \BibitemOpen
  \bibfield  {author} {\bibinfo {author} {\bibfnamefont {R.~J.}\ \bibnamefont
  {Stevens}}, \bibinfo {author} {\bibfnamefont {L.~A.}\ \bibnamefont
  {Mart{\'\i}nez-Tossas}},\ and\ \bibinfo {author} {\bibfnamefont
  {C.}~\bibnamefont {Meneveau}},\ }\bibfield  {title} {\bibinfo {title}
  {Comparison of wind farm large eddy simulations using actuator disk and
  actuator line models with wind tunnel experiments},\ }\href@noop {}
  {\bibfield  {journal} {\bibinfo  {journal} {Renewable energy}\ }\textbf
  {\bibinfo {volume} {116}},\ \bibinfo {pages} {470} (\bibinfo {year}
  {2018})}\BibitemShut {NoStop}%
\bibitem [{\citenamefont {Ribeiro}\ \emph {et~al.}(2020)\citenamefont
  {Ribeiro}, \citenamefont {Yeh},\ and\ \citenamefont
  {Taira}}]{Ribeiro2020Randomized}%
  \BibitemOpen
  \bibfield  {author} {\bibinfo {author} {\bibfnamefont {J.~H.~M.}\
  \bibnamefont {Ribeiro}}, \bibinfo {author} {\bibfnamefont {C.-A.}\
  \bibnamefont {Yeh}},\ and\ \bibinfo {author} {\bibfnamefont {K.}~\bibnamefont
  {Taira}},\ }\bibfield  {title} {\bibinfo {title} {Randomized resolvent
  analysis},\ }\href@noop {} {\bibfield  {journal} {\bibinfo  {journal}
  {Physical Review Fluids}\ }\textbf {\bibinfo {volume} {5}},\ \bibinfo {pages}
  {033902} (\bibinfo {year} {2020})}\BibitemShut {NoStop}%
\bibitem [{\citenamefont {Barthel}\ \emph {et~al.}(2022)\citenamefont
  {Barthel}, \citenamefont {Gomez},\ and\ \citenamefont
  {McKeon}}]{Barthel2022Variational}%
  \BibitemOpen
  \bibfield  {author} {\bibinfo {author} {\bibfnamefont {B.}~\bibnamefont
  {Barthel}}, \bibinfo {author} {\bibfnamefont {S.}~\bibnamefont {Gomez}},\
  and\ \bibinfo {author} {\bibfnamefont {B.~J.}\ \bibnamefont {McKeon}},\
  }\bibfield  {title} {\bibinfo {title} {Variational formulation of resolvent
  analysis},\ }\href@noop {} {\bibfield  {journal} {\bibinfo  {journal}
  {Physical Review Fluids}\ }\textbf {\bibinfo {volume} {7}},\ \bibinfo {pages}
  {013905} (\bibinfo {year} {2022})}\BibitemShut {NoStop}%
\bibitem [{\citenamefont {Hariharan}\ \emph {et~al.}(2021)\citenamefont
  {Hariharan}, \citenamefont {Kumar},\ and\ \citenamefont
  {Jovanovi{\'c}}}]{hariharan2021well}%
  \BibitemOpen
  \bibfield  {author} {\bibinfo {author} {\bibfnamefont {G.}~\bibnamefont
  {Hariharan}}, \bibinfo {author} {\bibfnamefont {S.}~\bibnamefont {Kumar}},\
  and\ \bibinfo {author} {\bibfnamefont {M.~R.}\ \bibnamefont
  {Jovanovi{\'c}}},\ }\bibfield  {title} {\bibinfo {title} {Well-conditioned
  ultraspherical and spectral integration methods for resolvent analysis of
  channel flows of newtonian and viscoelastic fluids},\ }\href@noop {}
  {\bibfield  {journal} {\bibinfo  {journal} {Journal of Computational
  Physics}\ }\textbf {\bibinfo {volume} {439}},\ \bibinfo {pages} {110241}
  (\bibinfo {year} {2021})}\BibitemShut {NoStop}%
\bibitem [{\citenamefont {Jin}\ \emph {et~al.}(2020)\citenamefont {Jin},
  \citenamefont {Illingworth},\ and\ \citenamefont
  {Sandberg}}]{jin2020feedback}%
  \BibitemOpen
  \bibfield  {author} {\bibinfo {author} {\bibfnamefont {B.}~\bibnamefont
  {Jin}}, \bibinfo {author} {\bibfnamefont {S.~J.}\ \bibnamefont
  {Illingworth}},\ and\ \bibinfo {author} {\bibfnamefont {R.~D.}\ \bibnamefont
  {Sandberg}},\ }\bibfield  {title} {\bibinfo {title} {Feedback control of
  vortex shedding using a resolvent-based modelling approach},\ }\href@noop {}
  {\bibfield  {journal} {\bibinfo  {journal} {Journal of Fluid Mechanics}\
  }\textbf {\bibinfo {volume} {897}},\ \bibinfo {pages} {A26} (\bibinfo {year}
  {2020})}\BibitemShut {NoStop}%
\bibitem [{\citenamefont {Batchelor}\ and\ \citenamefont
  {Gill}(1962)}]{batchelor1962analysis}%
  \BibitemOpen
  \bibfield  {author} {\bibinfo {author} {\bibfnamefont {G.}~\bibnamefont
  {Batchelor}}\ and\ \bibinfo {author} {\bibfnamefont {A.}~\bibnamefont
  {Gill}},\ }\bibfield  {title} {\bibinfo {title} {Analysis of the stability of
  axisymmetric jets},\ }\href@noop {} {\bibfield  {journal} {\bibinfo
  {journal} {Journal of fluid mechanics}\ }\textbf {\bibinfo {volume} {14}},\
  \bibinfo {pages} {529} (\bibinfo {year} {1962})}\BibitemShut {NoStop}%
\bibitem [{\citenamefont {Oberleithner}\ \emph {et~al.}(2014)\citenamefont
  {Oberleithner}, \citenamefont {Rukes},\ and\ \citenamefont
  {Soria}}]{oberleithner2014mean}%
  \BibitemOpen
  \bibfield  {author} {\bibinfo {author} {\bibfnamefont {K.}~\bibnamefont
  {Oberleithner}}, \bibinfo {author} {\bibfnamefont {L.}~\bibnamefont
  {Rukes}},\ and\ \bibinfo {author} {\bibfnamefont {J.}~\bibnamefont {Soria}},\
  }\bibfield  {title} {\bibinfo {title} {Mean flow stability analysis of
  oscillating jet experiments},\ }\href@noop {} {\bibfield  {journal} {\bibinfo
   {journal} {Journal of fluid mechanics}\ }\textbf {\bibinfo {volume} {757}},\
  \bibinfo {pages} {1} (\bibinfo {year} {2014})}\BibitemShut {NoStop}%
\bibitem [{\citenamefont {Matsushima}\ and\ \citenamefont
  {Marcus}(1995)}]{matsushima1995spectral}%
  \BibitemOpen
  \bibfield  {author} {\bibinfo {author} {\bibfnamefont {T.}~\bibnamefont
  {Matsushima}}\ and\ \bibinfo {author} {\bibfnamefont {P.}~\bibnamefont
  {Marcus}},\ }\bibfield  {title} {\bibinfo {title} {A spectral method for
  polar coordinates},\ }\href@noop {} {\bibfield  {journal} {\bibinfo
  {journal} {Journal of Computational Physics}\ }\textbf {\bibinfo {volume}
  {120}},\ \bibinfo {pages} {365} (\bibinfo {year} {1995})}\BibitemShut
  {NoStop}%
\bibitem [{\citenamefont {Smits}\ \emph {et~al.}(2011)\citenamefont {Smits},
  \citenamefont {McKeon},\ and\ \citenamefont {Marusic}}]{smits2011high}%
  \BibitemOpen
  \bibfield  {author} {\bibinfo {author} {\bibfnamefont {A.~J.}\ \bibnamefont
  {Smits}}, \bibinfo {author} {\bibfnamefont {B.~J.}\ \bibnamefont {McKeon}},\
  and\ \bibinfo {author} {\bibfnamefont {I.}~\bibnamefont {Marusic}},\
  }\bibfield  {title} {\bibinfo {title} {High--reynolds number wall
  turbulence},\ }\href@noop {} {\bibfield  {journal} {\bibinfo  {journal}
  {Annual Review of Fluid Mechanics}\ }\textbf {\bibinfo {volume} {43}},\
  \bibinfo {pages} {353} (\bibinfo {year} {2011})}\BibitemShut {NoStop}%
\bibitem [{\citenamefont {Crespo}\ \emph {et~al.}(1996)\citenamefont {Crespo},
  \citenamefont {Herna} \emph {et~al.}}]{crespo1996turbulence}%
  \BibitemOpen
  \bibfield  {author} {\bibinfo {author} {\bibfnamefont {A.}~\bibnamefont
  {Crespo}}, \bibinfo {author} {\bibfnamefont {J.}~\bibnamefont {Herna}}, \emph
  {et~al.},\ }\bibfield  {title} {\bibinfo {title} {Turbulence characteristics
  in wind-turbine wakes},\ }\href@noop {} {\bibfield  {journal} {\bibinfo
  {journal} {Journal of wind engineering and industrial aerodynamics}\ }\textbf
  {\bibinfo {volume} {61}},\ \bibinfo {pages} {71} (\bibinfo {year}
  {1996})}\BibitemShut {NoStop}%
\bibitem [{\citenamefont {Panofsky}\ and\ \citenamefont
  {Dutton}(1984)}]{panofsky1984atmospheric}%
  \BibitemOpen
  \bibfield  {author} {\bibinfo {author} {\bibfnamefont {H.~A.}\ \bibnamefont
  {Panofsky}}\ and\ \bibinfo {author} {\bibfnamefont {J.~A.}\ \bibnamefont
  {Dutton}},\ }\bibfield  {title} {\bibinfo {title} {Atmospheric turbulence.
  models and methods for engineering applications},\ }\href@noop {} {\bibfield
  {journal} {\bibinfo  {journal} {New York: Wiley}\ } (\bibinfo {year}
  {1984})}\BibitemShut {NoStop}%
\bibitem [{\citenamefont {Butler}\ and\ \citenamefont
  {Farrell}(1992)}]{butler1992three}%
  \BibitemOpen
  \bibfield  {author} {\bibinfo {author} {\bibfnamefont {K.~M.}\ \bibnamefont
  {Butler}}\ and\ \bibinfo {author} {\bibfnamefont {B.~F.}\ \bibnamefont
  {Farrell}},\ }\bibfield  {title} {\bibinfo {title} {Three-dimensional optimal
  perturbations in viscous shear flow},\ }\href@noop {} {\bibfield  {journal}
  {\bibinfo  {journal} {Physics of Fluids A: Fluid Dynamics}\ }\textbf
  {\bibinfo {volume} {4}},\ \bibinfo {pages} {1637} (\bibinfo {year}
  {1992})}\BibitemShut {NoStop}%
\bibitem [{\citenamefont {Pickering}\ \emph {et~al.}(2020)\citenamefont
  {Pickering}, \citenamefont {Rigas}, \citenamefont {Nogueira}, \citenamefont
  {Cavalieri}, \citenamefont {Schmidt},\ and\ \citenamefont
  {Colonius}}]{pickering2020lift}%
  \BibitemOpen
  \bibfield  {author} {\bibinfo {author} {\bibfnamefont {E.}~\bibnamefont
  {Pickering}}, \bibinfo {author} {\bibfnamefont {G.}~\bibnamefont {Rigas}},
  \bibinfo {author} {\bibfnamefont {P.~A.}\ \bibnamefont {Nogueira}}, \bibinfo
  {author} {\bibfnamefont {A.~V.}\ \bibnamefont {Cavalieri}}, \bibinfo {author}
  {\bibfnamefont {O.~T.}\ \bibnamefont {Schmidt}},\ and\ \bibinfo {author}
  {\bibfnamefont {T.}~\bibnamefont {Colonius}},\ }\bibfield  {title} {\bibinfo
  {title} {Lift-up, kelvin--helmholtz and orr mechanisms in turbulent jets},\
  }\href@noop {} {\bibfield  {journal} {\bibinfo  {journal} {Journal of Fluid
  Mechanics}\ }\textbf {\bibinfo {volume} {896}},\ \bibinfo {pages} {A2}
  (\bibinfo {year} {2020})}\BibitemShut {NoStop}%
\bibitem [{\citenamefont {Gupta}\ \emph {et~al.}(2021)\citenamefont {Gupta},
  \citenamefont {Madhusudanan}, \citenamefont {Wan}, \citenamefont
  {Illingworth},\ and\ \citenamefont {Juniper}}]{gupta2021linear}%
  \BibitemOpen
  \bibfield  {author} {\bibinfo {author} {\bibfnamefont {V.}~\bibnamefont
  {Gupta}}, \bibinfo {author} {\bibfnamefont {A.}~\bibnamefont {Madhusudanan}},
  \bibinfo {author} {\bibfnamefont {M.}~\bibnamefont {Wan}}, \bibinfo {author}
  {\bibfnamefont {S.~J.}\ \bibnamefont {Illingworth}},\ and\ \bibinfo {author}
  {\bibfnamefont {M.~P.}\ \bibnamefont {Juniper}},\ }\bibfield  {title}
  {\bibinfo {title} {Linear-model-based estimation in wall turbulence: improved
  stochastic forcing and eddy viscosity terms},\ }\href@noop {} {\bibfield
  {journal} {\bibinfo  {journal} {Journal of Fluid Mechanics}\ }\textbf
  {\bibinfo {volume} {925}},\ \bibinfo {pages} {A18} (\bibinfo {year}
  {2021})}\BibitemShut {NoStop}%
\bibitem [{\citenamefont {Wu}\ and\ \citenamefont
  {He}(2023)}]{wu2023composition}%
  \BibitemOpen
  \bibfield  {author} {\bibinfo {author} {\bibfnamefont {T.}~\bibnamefont
  {Wu}}\ and\ \bibinfo {author} {\bibfnamefont {G.}~\bibnamefont {He}},\
  }\bibfield  {title} {\bibinfo {title} {Composition of resolvents enhanced by
  random sweeping for large-scale structures in turbulent channel flows},\
  }\href@noop {} {\bibfield  {journal} {\bibinfo  {journal} {Journal of Fluid
  Mechanics}\ }\textbf {\bibinfo {volume} {956}},\ \bibinfo {pages} {A31}
  (\bibinfo {year} {2023})}\BibitemShut {NoStop}%
\bibitem [{\citenamefont {Madhusudanan}\ \emph {et~al.}(2022)\citenamefont
  {Madhusudanan}, \citenamefont {Illingworth}, \citenamefont {Marusic},\ and\
  \citenamefont {Chung}}]{madhusudanan2022navier}%
  \BibitemOpen
  \bibfield  {author} {\bibinfo {author} {\bibfnamefont {A.}~\bibnamefont
  {Madhusudanan}}, \bibinfo {author} {\bibfnamefont {S.~J.}\ \bibnamefont
  {Illingworth}}, \bibinfo {author} {\bibfnamefont {I.}~\bibnamefont
  {Marusic}},\ and\ \bibinfo {author} {\bibfnamefont {D.}~\bibnamefont
  {Chung}},\ }\bibfield  {title} {\bibinfo {title} {Navier-stokes--based linear
  model for unstably stratified turbulent channel flows},\ }\href@noop {}
  {\bibfield  {journal} {\bibinfo  {journal} {Physical Review Fluids}\ }\textbf
  {\bibinfo {volume} {7}},\ \bibinfo {pages} {044601} (\bibinfo {year}
  {2022})}\BibitemShut {NoStop}%
\end{thebibliography}%

%

\end{document}